\documentclass{aa}  

\usepackage{graphicx}
\usepackage{epsfig}
\usepackage{pdflscape}
\usepackage{afterpage}
\usepackage{txfonts}
\begin{document} 

   \title{Predicted microlensing events from analysis of \\
          \textit{Gaia} Data Release 2}

   \author{D.M. Bramich
          \inst{1}
          }

   \institute{New York University Abu Dhabi, PO Box 129188, Saadiyat Island, Abu Dhabi, UAE\\
              \email{dan.bramich@hotmail.co.uk}
             }

   \date{Received May 26, 2018; accepted ???, 2018}

 
  \abstract
   {
   Astrometric microlensing can be used to make precise measurements of the masses of lens stars that are independent
   of their assumed internal physics. Such direct mass measurements, obtained purely by observing the gravitational
   effects of the stars on external objects, are crucial for validating theoretical stellar models. Specifically,
   astrometric microlensing provides a channel to direct mass measurements of single stars for which so few measurements exist.
   Microlensing events that also exhibit a detectable photometric signature provide even stronger lens mass constraints.
   }
   {
   To use the astrometric solutions and photometric measurements of $\sim$1.7~billion stars provided by \textit{Gaia}
   Data Release 2 (GDR2) to predict microlensing events during the nominal \textit{Gaia} mission and beyond.
   This will enable astronomers to observe the entirety of each event, including the peak, with appropriate observing resources.
   The data collected will allow precise lens mass measurements for white dwarfs and low-mass main sequence stars
   (K and M dwarfs) helping to constrain stellar evolutionary models.
   }
   {
   I search for source-lens pairs in GDR2 that could potentially lead to microlensing events between 25th July 2014 and
   25th July 2026. I estimate the lens masses using GDR2 photometry and parallaxes, and appropriate model stellar isochrones.
   Combined with the source and lens parallax measurements from GDR2, this allows the Einstein ring radius
   to be computed for each source-lens pair. By considering the source and lens paths on the sky, I calculate the microlensing signals
   that are to be expected.
   }
   {
   I present a list of 76 predicted microlensing events. Nine and five astrometric events will be caused by
   the white dwarf stars LAWD~37 and Stein~2051~B, respectively. A further nine events will exhibit detectable photometric and astrometric signatures.
   Of the remaining events, ten will exhibit astrometric signals with peak amplitudes above 0.5~mas, while the rest are
   low-amplitude astrometric events with peak amplitudes between 0.131 and 0.5~mas. Five and two events will reach their peaks during 2018 and 2019,
   respectively. Five of the photometric events have the potential to evolve into high-magnification events, which may also probe for
   planetary companions to the lenses.
   }
   {}


   \maketitle

%
\section{Introduction}
\label{sec:intro}

Gravitational lensing involves the bending of light rays from a source by a massive
object (referred to as the lens; \citealt{ein1915}). When a single compact lens object passes close enough to the line of sight
between the source and the observer, then the observer typically views two distorted and magnified images of
the background source (\citealt{ein1936}; \citealt{lie1964}; \citealt{ref1964}).
Within our own Galaxy, sources and lenses are usually stars and the lensed
images are separated by less than a few milliarcseconds\footnote{For a source star at 8~kpc and a solar-mass lens
star at 4~kpc, the Einstein radius is $\sim$1~mas.} (mas). This is unresolvable with currently
available telescopes (both ground and space) and the effect is referred to as microlensing (\citealt{pet1981}; \citealt{pac1986}).

As the source and lens move relative to each other, the strength of the lensing effect changes,
leading to observable deviations in source brightness (photometric microlensing; e.g. \citealt{bea2006}) and centroid position
(astrometric microlensing; e.g. \citealt{sah2017}). For a lens that is too faint to be detected,
observation of both the photometric and astrometric microlensing signals allows important
parameter degeneracies to be broken (\citealt{hog1995}; \citealt{miy1995}; \citealt{wal1995}),
although no microlensing event has yet been detected via both channels to date. For bright lenses where
both the source and the lens have known distances via trigonometric parallax measurements, the astrometric lensing signal
alone enables the mass of the lens to be determined.
This has recently been achieved for the first time ever by \citet{sah2017} for the white dwarf Stein~2051~B.
They used \textit{Hubble Space Telescope} (\textit{HST}) observations acquired at eight epochs over two years to measure a $\sim$3~mas deflection in the position
of an 18.3~mag ($V$ band) background source star. Combined with the parallax measurements for the source and the lens,
the mass of Stein~2051~B was measured as $\sim$0.675~$M_{\sun}$ with uncertainty $\sim$8\%, which provided
a new sorely-needed datum for comparison with white dwarf evolutionary models.

The validation of theoretical models of stars requires mass measurements that are independent of the assumed
internal physics. Direct mass measurements, obtained purely by observing the gravitational effects of
stars on external objects, fulfill these requirements. However, most directly measured stellar masses come
from observations of the orbital motion of binary stars (\citealt{tor2010}), and stars in binary systems evolve differently to
single stars. Hence, direct mass measurements of single stars are highly desirable and exceptionally important. Astrometric
microlensing provides a powerful technique for the direct determination of masses of single stars proven to yield uncertainties
below $\sim$10\%, and that has the potential to achieve uncertainties of $\sim$1\%. This compares favourably with the best uncertainties in mass
estimates for stars in binary systems ($\sim$1-3\%) and from asteroseismology ($\sim$1-10\%; \citealt{cha2014}; \citealt{sil2017}).

Given that microlensing events are intrinsically rare occurrences that depend on chance stellar alignments, predicting when
and where they will occur is highly advantageous for the collection of data throughout an event. The first attempt
at predicting a microlensing event was made by \citet{fei1966}. Frustratingly for
Feibelman\footnote{In \citet{fei1986}, he concludes: ``It is hoped that this 20-year exercise in frustration will
encourage others to conduct systematic searches ... for stars with large proper motions that eventually may eclipse
a background star and give rise to the elusive gravitational lens effect''. My work in this paper is a contemporary realisation of this
sentiment.}, subsequent observations showed that the potential lens star 40~Eridani~A would not pass close enough to the source star
to yield a detectable signal (\citealt{fei1986}). Nowadays, thousands of
photometric microlensing events towards the Galactic bulge are routinely detected each year by dedicated surveys
(Optical Gravitational Lensing Experiment - OGLE - \citealt{uda2003}; Microlensing Observations in Astrophysics - MOA - \citealt{bon2001})
which alert the astronomical community to the events as early as possible to enable
follow-up observations in the search for extrasolar planets (e.g. RoboNet-II - \citealt{tsa2009}).
Unfortunately, due to the small signals involved, there are no dedicated ground-based surveys for discovering astrometric
microlensing events, while from space, targeted efforts are ongoing (\citealt{kai2017}).

The \textit{Gaia} satellite (see Section~\ref{sec:gaia}), with its unprecedented combination of all-sky coverage, 
sample volume (depth), and astrometric precision and accuracy, is predicted to serendipitously
detect thousands of astrometric microlensing events during
its five-year mission (\citealt{dom2000}; \citealt{bel2002}). However, the prediction of exactly which stars will undergo
microlensing deviations is only possible for the subset of events
with bright lenses\footnote{It is interesting to note that the idea of predicting microlensing events can be traced back at least
as far as \citet{ref1964} who concluded that ``By comparing photographs of the sky taken at different times, the angular
velocity of a great number of stars can be determined, and passages may be predicted.''}.
This task was attempted by Proft, Demleitner \& Wambsganss (2011; hereafter P11)
using a patchwork of the best stellar proper motion catalogues available at the time. 
Unfortunately, the precision of the stellar positions and proper motions in the catalogues that were used,
which are limited by the ground-based observations from which they are derived, is not good enough
for predicting microlensing events with a high certainty. Furthermore, parallax measurements,
and therefore distances, were unavailable for the majority of the candidate lens stars listed in P11,
substantially increasing the uncertainty in any supposed microlensing geometry. The combined effect
of these unavoidable limitations resulted in highly uncertain estimates of the timings and amplitudes
of the astrometric signals for the candidate events (see Table~2 in P11).
Finally, due to the fact that the \textit{Gaia} launch was delayed, and since the mission is also likely to be extended,
a reanalysis of the predicted events during the revised observational window is required anyway.

A few days before \textit{Gaia} Data Release 2 (\citealt{bro2018}; hereafter GDR2; Section~\ref{sec:gaia})
and the submission of this paper, McGill at al. (2018; hereafter M18)
reported on a predicted astrometric microlensing event that will be caused by the white dwarf LAWD~37 (WD~1142$-$645).
M18 predict that the event will exhibit a peak astrometric signal on 11th November 2019 ($\pm$4~d), which is just after the end of
the \textit{Gaia} nominal five-year mission. It was the only candidate event
discovered from an analysis of the Tycho-\textit{Gaia} Astrometric Solution
and \textit{Gaia} Data Release 1 catalogues (\citealt{lin2016}). In fact, it turns out that LAWD~37 will
cause nine microlensing events during the time baseline of any extended \textit{Gaia} mission (see Section~\ref{sec:lawd37}).

In this paper, I use GDR2 as a precise and accurate astrometric and photometric catalogue of stars to predict microlensing
events with detectable signals during the (extended) lifetime of the \textit{Gaia} satellite.
In Section~\ref{sec:microastro}, I review the essential aspects of microlensing relevant to this paper. In Section~\ref{sec:methods},
I describe the methods I use to identify source-lens pairs from GDR2 that could potentially lead to detectable microlensing events.
I estimate lens masses and search for microlensing events in Section~\ref{sec:events}, and I present
my results in Section~\ref{sec:results}.
I provide a brief summary and conclusions in Section~\ref{sec:conc}.

\section{The \textit{Gaia} Satellite And Data Release 2}
\label{sec:gaia}

The \textit{Gaia} satellite (Prusti et al. 2016; hereafter P16),
launched on 19th December 2013, aims to measure the three-dimensional spatial and velocity
distribution of stars over a large fraction of the volume of the Galaxy, and to determine their properties such as
effective temperature and surface gravity. These measurements for $\sim$10$^{9}$~stars will enable a vastly
improved understanding of (i) the structure, dynamics, and evolution of our Galaxy, including its star formation history,
(ii) stellar physics and evolution, (iii) binary and multiple stars, and (iv) stellar variability and the distance scale
(see \citealt{per2001}). This list of science applications is not exhaustive. Science observations for the nominal five-year
mission started on 25th July 2014. However, it has already been announced\footnote{\url{https://www.cosmos.esa.int/web/gaia/news}}
that \textit{Gaia} operations will likely continue until at least the end of 2020, although the principal science mission
lifetime is limited to 10$\pm$1~years by the amount of fuel available for the fine attitude control thrusters (P16).

\textit{Gaia} operates at the second Lagrange point (L$_{2}$) of the Sun-Earth-Moon system, and it scans the entire
sky using uniform revolving scanning with two identical telescopes separated by a constant angle of 106\fdg5 on the sky
along the scanning circle. The spin rate around the spacecraft spin axis is one revolution every 6 hours
(or equivalently 60\,\arcsec~s$^{-1}$), and the spin axis precesses slowly around the Sun
(5.8 revolutions per year) at a fixed solar-aspect angle of 45\degr.
Hence the scanning law provides at least six distinct epochs of observation per year for any object in the sky
(and two observations at each epoch), although the number of times any particular object is observed depends on its
position in the sky (especially the ecliptic latitude). The sky-averaged end-of-mission number of observations is $\sim$80.
Objects scan across the 106 charge-coupled device (CCD) detectors in the shared focal plane of the two telescopes while the
CCDs are operated in time-delayed integration mode. Pixel windows around objects of interest are kept and binning
is performed in the across-scan direction.

\textit{Gaia} performs absolute astrometry in the International Celestial Reference System (ICRS; \citealt{ari1995})
for objects between 3~mag and 21~mag. A single observation of an object yields an astrometric measurement derived
from the transit of the object over both a sky mapper CCD and nine astrometric field CCDs. The along-scan positional
error is much smaller than the across-scan positional error (by a factor of $\sim$7-70; Section~\ref{sec:astroprec}).
However, the varying scanning angles executed by \textit{Gaia} ensure that precise two-dimensional astrometric information
can be recovered from what are effectively one-dimensional measurements. The standard errors on the positional measurement
for a single observation in the along- and across-scan directions are typically $\sim$0.03 and 0.32~mas at 10 mag, $\sim$0.12 and 
5.4~mas at 15 mag, and $\sim$3.1 and 200~mas at 20 mag, respectively.

The satellite also performs white-light photometry ($G$ passband, 330-1050~nm), and low-resolution
spectroscopy which integrates to magnitudes in broad blue and red passbands $G_{\mbox{\scriptsize BP}}$ (330-680~nm)
and $G_{\mbox{\scriptsize RP}}$ (630-1050~nm), respectively (\citealt{car2016}; \citealt{eva2018}). The standard error on the $G$ magnitude
for a single observation is $\sim$0.4~mmag for $G<12$~mag, $\sim$1.7~mmag at $G=15$~mag, and $\sim$31~mmag at $G=20$~mag
(Section~\ref{sec:photprec}).

The astrometric crowding limit for \textit{Gaia} is $\sim$1,050,000 objects per deg$^{2}$ for $G$-band photometry, 
and $\sim$750,000 objects per deg$^{2}$ for $G_{\mbox{\scriptsize BP}}$- and $G_{\mbox{\scriptsize RP}}$-band photometry (P16). Stars
with equal brightness can be resolved down to $\sim$100~mas since the along-scan point-spread function (PSF) full-width at half-maximum
(FWHM) has values that fall mostly between $\sim$90-108~mas (median value $\sim$103~mas; see Figure~9 of \citealt{fab2016}).
Larger separations are required to resolve stars with brightness differences.

GDR2 corresponds to all science data obtained by \textit{Gaia} between 25th July 2014 (10:30~UTC) and 23rd May 2016 (11:35~UTC).
It contains 1,692,919,135 objects of which 1,331,909,727 have 5-parameter astrometric solutions (position, proper motion, and
parallax), and 361,009,408 have 2-parameter astrometric solutions (position only). The astrometric solutions are based
on \textit{Gaia} data alone. All objects have a mean $G$ magnitude measurement,
and $\sim$82\% of objects also have mean $G_{\mbox{\scriptsize BP}}$ and $G_{\mbox{\scriptsize RP}}$ photometric measurements.
Typical uncertainties in the proper motions range from $\sim$0.05~mas~yr$^{-1}$ for $G<14$~mag to $\sim$1.2~mas~yr$^{-1}$
at $G=20$~mag. For parallaxes and positions at the reference epoch J2015.5,
the uncertainties range from $\sim$0.04~mas for $G<14$~mag to $\sim$0.7~mas at $G=20$~mag (\citealt{lin2018}).

Some of the limitations of GDR2 have a negative impact on predicting the microlensing events that \textit{Gaia} will
observe. GDR2 has a faint limit that varies across the sky due mainly to the inhomogeneous scanning coverage over the
corresponding time period and the spatial variations in stellar density. This adversely affects the number of potential
source stars available to my study. GDR2 also reports that a fraction of stars brighter than $G=7$~mag are missing
from the catalogue and that the completeness near bright objects is not yet optimal, which affects both potential source
and lens stars in my analysis. Finally, $\sim$17\% of stars with proper motions greater than $\sim$0.6~arcsec~yr$^{-1}$
are missing from GDR2 (\citealt{bro2018}), which adversely affects stars that
are potentially some of the best lens candidates. Hence the list of predicted microlensing events with bright
lenses set forth in this paper is likely to be somewhat incomplete when compared in hindsight to the actual list of 
such events detected by \textit{Gaia}.

\section{Microlensing Essentials}
\label{sec:microastro}

The theoretical treatment of microlensing has been explored
in detail throughout the relevant literature (e.g. \citealt{pac1996}).
Consequently, I limit myself to summarising and developing the important concepts and
equations directly relevant to this paper. Specifically I follow Dominik \& Sahu (2000; hereafter D00; see also
references therein).

\subsection{Point-Source Point-Lens Lensing}
\label{sec:pspl}

\begin{figure*}
\centering
\epsfig{file=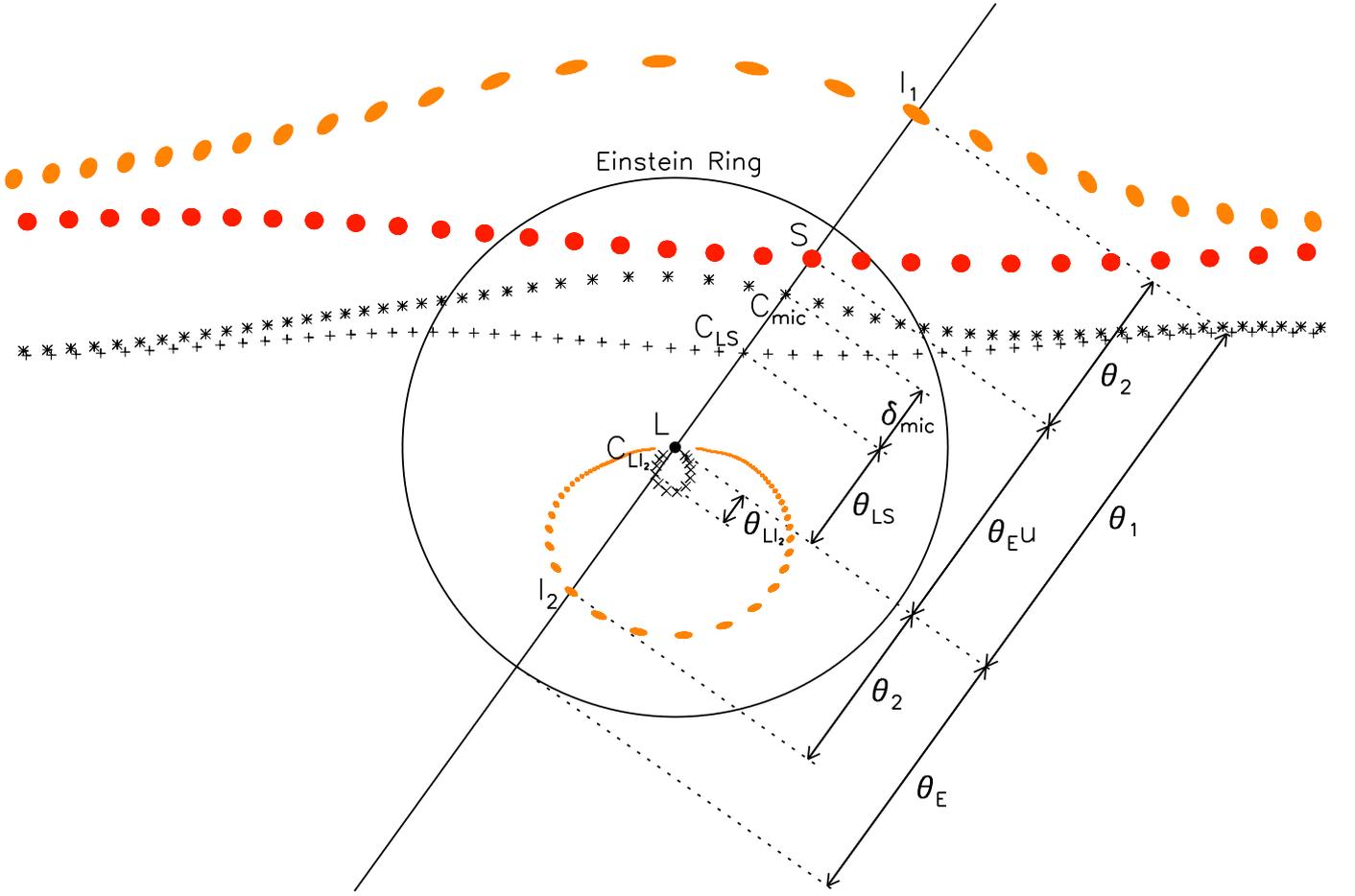,angle=0.0,width=\linewidth}
\caption{Configuration of a microlensing event as seen by the observer, who does not resolve any of the components.
         The filled red circles represent the source S at equal time intervals as it moves relative to the
         lens L (black circle). The relative motion of the source consists of uniform proper motion in a straight line combined with
         annual parallax, which is why the source appears to move in a non-linear fashion.
         The major and minor images, I$_{1}$ and I$_{2}$, respectively,
         of the source are plotted as orange ellipses. Although S, L, I$_{1}$ and I$_{2}$ are all plotted with a finite size
         for illustrative purposes, the analysis in this paper ignores finite-size effects. At any single instant,
         I$_{1}$, S, L and I$_{2}$ form a straight line on the sky along with the centroids C$_{\mbox{\scriptsize mic}}$ (asterisks),
         C$_{\mbox{\scriptsize LS}}$ (plus signs) and C$_{\mbox{\scriptsize LI}_{2}}$ (crosses). 
         The centroids C$_{\mbox{\scriptsize mic}}$, C$_{\mbox{\scriptsize LS}}$ and C$_{\mbox{\scriptsize LI}_{2}}$
         have been plotted for $f_{\mbox{\scriptsize L}} / f_{\mbox{\scriptsize S}} = 1$. Angular distances defined in
         the text are marked in the diagram.
         \label{fig:microlensing}}
\end{figure*}

Let a (point-like) source S and (point-like) lens L be at distances $D_{\mbox{\scriptsize S}}$ and $D_{\mbox{\scriptsize L}}$, respectively, from an observer
such that $0<D_{\mbox{\scriptsize L}}<D_{\mbox{\scriptsize S}}$. Given angular position vectors $\vec{\phi}_{\mbox{\scriptsize S}}$ and
$\vec{\phi}_{\mbox{\scriptsize L}}$ on the celestial sphere for the source and lens, respectively, one can define the dimensionless
distance vector:
\begin{equation}
\vec{u} = \frac{ \vec{\phi}_{\mbox{\scriptsize S}} - \vec{\phi}_{\mbox{\scriptsize L}} }{ \theta_{\mbox{\scriptsize E}} }
\label{eqn:uvec}
\end{equation}
where $\theta_{\mbox{\scriptsize E}}$ is the Einstein radius:
\begin{equation}
\theta_{\mbox{\scriptsize E}} = \sqrt{ \frac{4\,G M}{c^{2}} \left( \frac{1}{ D_{\mbox{\scriptsize L}} } - \frac{1}{ D_{\mbox{\scriptsize S}} } \right) }
\label{eqn:einsteinradius}
\end{equation}
The mass $M$ is the mass of the lens.
The major and minor images, I$_{1}$ and I$_{2}$, respectively, of the source as seen by the observer
are located on either side of the lens along the straight line
I$_{2}$-L-S-I$_{1}$ in the sky plane (Figure~\ref{fig:microlensing}).
I$_{1}$ and I$_{2}$ lie outside and inside, respectively, of the Einstein ring around the lens.
They are separated by an angular distance on the sky of:
\begin{equation}
\theta_{\mbox{\scriptsize sep}} = \theta_{1} + \theta_{2} = \theta_{\mbox{\scriptsize E}} \sqrt{ u^{2} + 4 }
                                \ge \theta_{\mbox{\scriptsize E}} \, u
\label{eqn:thetasep}
\end{equation}
where $u = |\vec{u}|$.
The angular distances $\theta_{1}$ and $\theta_{2}$ on the sky corresponding to L-I$_{1}$ and L-I$_{2}$, respectively, are given by:
\begin{equation}
\theta_{1} = \frac{ \theta_{\mbox{\scriptsize E}} }{2} \left( \sqrt{ u^{2} + 4 } + u \right) \ge \theta_{\mbox{\scriptsize E}} \, u
\label{eqn:theta1}
\end{equation}
\begin{equation}
\theta_{2} = \theta_{1} - \theta_{\mbox{\scriptsize E}} \, u \ge 0
\label{eqn:theta2}
\end{equation}
The flux ratios $A_{1}$ and $A_{2}$, relative to the source, of the images I$_{1}$ and I$_{2}$, respectively,
are:
\begin{equation}
A_{1} = \frac{ u^{2} + 2 }{ 2 u \sqrt{u^{2} + 4} } + \frac{1}{2} \ge 1
\label{eqn:a1}
\end{equation}
\begin{equation}
A_{2} = A_{1} - 1 \ge 0
\label{eqn:a2}
\end{equation}

\subsection{Microlensing}
\label{sec:microlensing}

In microlensing, the source images I$_{1}$ and I$_{2}$, and the lens L, are all unresolved. Furthermore, it is possible that
the lens is luminous\footnote{However, here I assume that any ``third light'' from unrelated objects that happen to be
blended with the microlensing system is negligible.}. Adopting $f_{\mbox{\scriptsize S}}$ and $f_{\mbox{\scriptsize L}}$
as the source and lens fluxes, respectively, 
then one may write the observed overall magnification (brightening) of the microlensing system as:
\begin{equation}
\begin{aligned}
A & = \frac{ A_{1} \, f_{\mbox{\scriptsize S}} + A_{2} \, f_{\mbox{\scriptsize S}} + f_{\mbox{\scriptsize L}} }
         { f_{\mbox{\scriptsize S}} + f_{\mbox{\scriptsize L}} } \\
  & = \frac{ u^{2} + 2 + ( f_{\mbox{\scriptsize L}} / f_{\mbox{\scriptsize S}} ) \, u \sqrt{u^{2} + 4} }
           { ( 1 + f_{\mbox{\scriptsize L}} / f_{\mbox{\scriptsize S}} ) \, u \sqrt{u^{2} + 4} }
    \ge 1 \\
\label{eqn:magnification}
\end{aligned}
\end{equation}
For a non-luminous (dark) lens, one has $f_{\mbox{\scriptsize L}} = 0$, which yields the standard result:
\begin{equation}
A = \frac{ u^{2} + 2 }{ u \sqrt{u^{2} + 4} } \ge 1
\label{eqn:magnificationdarklens}
\end{equation}

Now consider the observed centroid C$_{\mbox{\scriptsize mic}}$ of the microlensing system (i.e. the centroid of I$_{1}$, I$_{2}$ and L).
It lies on the line L-I$_{1}$ (Figure~\ref{fig:microlensing}) at an angular distance from L of:
\begin{equation}
\begin{aligned}
\theta_{\mbox{\scriptsize mic}} & = \frac{ A_{1} \, \theta_{1} - A_{2} \, \theta_{2} }
                                         { A_{1} + A_{2} + f_{\mbox{\scriptsize L}} / f_{\mbox{\scriptsize S}} } \\
                                & = \theta_{\mbox{\scriptsize E}} \, u
                                    \left[ \frac{ u^{2} + 3 }
                                                { u^{2} + 2 + ( f_{\mbox{\scriptsize L}} / f_{\mbox{\scriptsize S}} ) \, u \sqrt{u^{2} + 4} } \right]
                                  \ge 0 \\
\label{eqn:centroidmic}
\end{aligned}
\end{equation}
In the absence of lensing effects, the observed centroid C$_{\mbox{\scriptsize LS}}$ of S and L,
which also lies on the line L-I$_{1}$, would be at an angular distance from L of:
\begin{equation}
\theta_{\mbox{\scriptsize LS}} = \frac{ \theta_{\mbox{\scriptsize E}} \, u }{ 1 + f_{\mbox{\scriptsize L}} / f_{\mbox{\scriptsize S}} } \ge 0
\label{eqn:centroidLS}
\end{equation}
which is \textit{only coincident with the source position when $f_{\mbox{\scriptsize L}} = 0$}.
The centroid shift due to microlensing may then be derived as:
\begin{equation}
\begin{aligned}
\delta_{\mbox{\scriptsize mic}} & = \theta_{\mbox{\scriptsize mic}} - \theta_{\mbox{\scriptsize LS}} \\
                                & = \frac{ \theta_{\mbox{\scriptsize E}} \, u }{ 1 + f_{\mbox{\scriptsize L}} / f_{\mbox{\scriptsize S}} }
                                    \left[ \frac{ 1 + ( f_{\mbox{\scriptsize L}} / f_{\mbox{\scriptsize S}} ) \,
                                                      \left( u^{2} + 3 - u \sqrt{u^{2} + 4} \right) }
                                                { u^{2} + 2 + ( f_{\mbox{\scriptsize L}} / f_{\mbox{\scriptsize S}} ) \, u \sqrt{u^{2} + 4} } \right]
                                  \ge 0 \\
\label{eqn:centroidshift}
\end{aligned}
\end{equation}
For a dark lens, one has $f_{\mbox{\scriptsize L}} = 0$, which yields the well-known result:
\begin{equation}
\delta_{\mbox{\scriptsize mic}} = \theta_{\mbox{\scriptsize E}} \left( \frac{u}{u^{2} + 2} \right) \ge 0
\label{eqn:centroidshiftdarklens}
\end{equation}

It is worth noting that incorrect results for $\delta_{\mbox{\scriptsize mic}}$ in Equation~\ref{eqn:centroidshift} have been derived
in the literature by calculating the centroid shift relative to the source position (e.g. \citealt{bod1998}) or by
failing to take into account the source image splitting (e.g. \citealt{gol1998a}, \citealt{gol1998b}). The only
authors to have presented the correct expression for $\delta_{\mbox{\scriptsize mic}}$ are
D00 and \citet{lee2010}. Unfortunately, one also finds that many derivations in the literature of $\delta_{\mbox{\scriptsize mic}}$
for the special case where $f_{\mbox{\scriptsize L}} = 0$ (Equation~\ref{eqn:centroidshiftdarklens}) are accompanied by
the potentially misleading statement that the centroid shift is being calculated relative to the source position
as opposed to the observed source-lens centroid (e.g. \citealt{gou2001}, \citealt{han2002}, \citealt{nuc2016}).

\subsection{Partially-Resolved Microlensing}
\label{sec:partialmicrolensing}

With the resolution and precision of \textit{Gaia}, it is possible that microlensing signals
are detectable even when $\theta_{\mbox{\scriptsize E}} \, u$ is large enough for L and I$_{1}$ to be resolved.
In this case, L and I$_{2}$ remain blended, and the ratio $A_{\mbox{\scriptsize LI}_{2}}$ of the blend flux
relative to the lens flux is given by:
\begin{equation}
\begin{aligned}
A_{\mbox{\scriptsize LI}_{2}} & = \frac{ A_{2} \, f_{\mbox{\scriptsize S}} + f_{\mbox{\scriptsize L}} }{ f_{\mbox{\scriptsize L}} } \\
                              & = \frac{ u^{2} + 2 + (2 ( f_{\mbox{\scriptsize L}} / f_{\mbox{\scriptsize S}} ) - 1)\, u \sqrt{u^{2} + 4} }
                                       { 2 ( f_{\mbox{\scriptsize L}} / f_{\mbox{\scriptsize S}} )\, u \sqrt{u^{2} + 4} }
                                  \ge 1 \\
\label{eqn:ALI2}
\end{aligned}
\end{equation}
Furthermore, the observed centroid C$_{\mbox{\scriptsize LI}_{2}}$ of L and I$_{2}$ lies
on the line L-I$_{2}$ (Figure~\ref{fig:microlensing}) at an angular distance from L of:
\begin{equation}
\begin{aligned}
\theta_{\mbox{\scriptsize LI}_{2}} & = \frac{ A_{2} \, \theta_{2} }{ A_{2} + f_{\mbox{\scriptsize L}} / f_{\mbox{\scriptsize S}} } \\
                                   & = \theta_{\mbox{\scriptsize E}}
                                       \left[ \frac{ (u^{2} + 1) \left( \sqrt{u^{2} + 4} - u \right) - 2u }
                                                   { u^{2} + 2 + (2 ( f_{\mbox{\scriptsize L}} / f_{\mbox{\scriptsize S}} ) - 1)\, u \sqrt{u^{2} + 4} } \right]
                                       \ge 0 \\
\label{eqn:centroidLI2}
\end{aligned}
\end{equation}
For a dark lens, one has $f_{\mbox{\scriptsize L}} = 0$ and $\theta_{\mbox{\scriptsize LI}_{2}} = \theta_{2}$ while $A_{\mbox{\scriptsize LI}_{2}}$
is ill-defined.

I refer to this situation as \textit{partially-resolved microlensing}\footnote{Alternatively,
this could also be referred to as \textit{partially-unresolved lensing}.}
because lensing effects are detectable while some, but not all, of the components
L, I$_{1}$ and I$_{2}$ are resolved.
In this regime, for a luminous lens ($f_{\mbox{\scriptsize L}}>0$), the lensing has two observable photometric and astrometric effects.
Firstly, the major source image I$_{1}$ is magnified by $A_{1}$ relative to the source flux and it is shifted from the nominal source
position S by an angular distance of $\theta_{2}$. Secondly, the blend of L and I$_{2}$ is magnified by $A_{\mbox{\scriptsize LI}_{2}}$
relative to the lens flux and the centroid C$_{\mbox{\scriptsize LI}_{2}}$ is shifted from the nominal lens position L by an angular distance
of $\theta_{\mbox{\scriptsize LI}_{2}}$ along the line L-I$_{2}$.

\subsection{Behaviour Of The Observed Signals}
\label{sec:microlensingbehaviour}

\begin{figure}
\centering
\epsfig{file=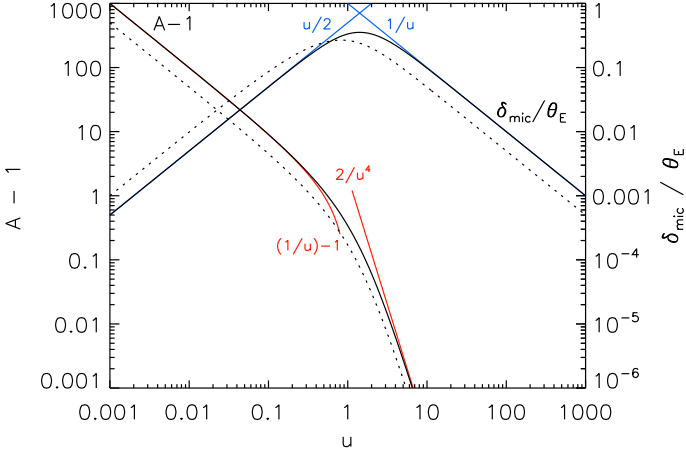,angle=0.0,width=\linewidth}
\caption{Variation of the magnification above baseline $A - 1$ and the normalised centroid shift
         $\delta_{\mbox{\scriptsize mic}} / \theta_{\mbox{\scriptsize E}}$
         during a microlensing event as a function of the normalised source-lens separation $u$. Continuous and dotted
         black curves correspond to the cases of a dark lens ($f_{\mbox{\scriptsize L}} = 0$) and a luminous lens
         with $f_{\mbox{\scriptsize L}} / f_{\mbox{\scriptsize S}} = 1$, respectively. The asymptotic behaviour of the photometric
         and astrometric signals for the specific case of a dark lens is plotted using red and blue curves, respectively, and these
         asymptotic curves are labelled with their functional form.
         \label{fig:comparison}}
\end{figure}

Figure~\ref{fig:comparison} shows the variation of the photometric and astrometric signals during a microlensing event as a
function of the normalised source-lens separation $u$. A microlensing event can potentially reach magnifications above $A \ga 1000$
for very small values of $u \la 0.001$, and it is only finite-source effects (\citealt{wit1994}) that limit the peak magnification
(e.g. \citealt{don2006}). In contrast, the normalised
centroid shift $\delta_{\mbox{\scriptsize mic}} / \theta_{\mbox{\scriptsize E}}$ reaches a maximum
near $u \approx 1$ and it tends to zero for very small values of $u$. Both $A-1$ and
$\delta_{\mbox{\scriptsize mic}} / \theta_{\mbox{\scriptsize E}}$ tend to zero for very large values of $u$, albeit
at considerably different rates. Curves are plotted in Figure~\ref{fig:comparison} for a dark lens (continuous black lines) and for
a luminous lens with $f_{\mbox{\scriptsize L}} / f_{\mbox{\scriptsize S}} = 1$ (dotted black lines). The effect of a
luminous lens is to decrease the amplitude of both the photometric and astrometric signals. Furthermore, the
maximum centroid shift occurs at a smaller value of $u$ for a luminous lens.

More formally, the asymptotic behaviour of $A$ and $\delta_{\mbox{\scriptsize mic}}$ for $u \ll 1$ is:
\begin{equation}
A \sim \frac{1}{(1 + f_{\mbox{\scriptsize L}} / f_{\mbox{\scriptsize S}}) \, u}
\label{eqn:magsmallu}
\end{equation}
\begin{equation}
\delta_{\mbox{\scriptsize mic}} \sim \frac{ \theta_{\mbox{\scriptsize E}} \, u }{2}
                                     \left[ \frac{ 1 + 3 f_{\mbox{\scriptsize L}} / f_{\mbox{\scriptsize S}} }
                                                 { 1 + f_{\mbox{\scriptsize L}} / f_{\mbox{\scriptsize S}} } \right]
\label{eqn:deltasmallu}
\end{equation}
For $u \gg 1$, the asymptotic behaviour is:
\begin{equation}
A \sim 1 + \frac{2}{(1 + f_{\mbox{\scriptsize L}} / f_{\mbox{\scriptsize S}})\, u^{4}}
\label{eqn:maglargeu}
\end{equation}
\begin{equation}
\delta_{\mbox{\scriptsize mic}} \sim \frac{\theta_{\mbox{\scriptsize E}}}{(1 + f_{\mbox{\scriptsize L}} / f_{\mbox{\scriptsize S}}) \, u}
\label{eqn:deltalargeu}
\end{equation}
It is useful to note from these equations that as $u$ increases from $\sqrt{2}$,
the magnification above baseline $A - 1$ decreases at a much faster rate than the centroid shift $\delta_{\mbox{\scriptsize mic}}$.

For a dark lens, the maximum centroid shift occurs when $u = \sqrt{2} \approx 1.414$ with the value
$\delta_{\mbox{\scriptsize mic}} \approx 0.354 \, \theta_{\mbox{\scriptsize E}}$. For a luminous
lens with $f_{\mbox{\scriptsize L}} / f_{\mbox{\scriptsize S}} = 1$, the maximum centroid shift occurs when
$u \approx 0.827$ with the value $\delta_{\mbox{\scriptsize mic}} \approx 0.267 \, \theta_{\mbox{\scriptsize E}}$.
In general, computing the value of $u$ that corresponds to the maximum centroid shift for a luminous lens requires
the solution of a high-degree polynomial.

In the partially-resolved microlensing regime, the most relevant asymptotic behaviour for the
two photometric and astrometric signals is for $u \gg 1$ where:
\begin{equation}
A_{1} \sim 1 + \frac{1}{u^{4}}
\label{eqn:A2largeu}
\end{equation}
\begin{equation}
\theta_{2} \sim \frac{\theta_{\mbox{\scriptsize E}}}{u}
\label{eqn:theta2largeu}
\end{equation}
\begin{equation}
A_{\mbox{\scriptsize LI}_{2}} \sim 1 + \frac{1}{(f_{\mbox{\scriptsize L}} / f_{\mbox{\scriptsize S}}) \, u^{4}}
\label{eqn:ALI2largeu}
\end{equation}
\begin{equation}
\theta_{\mbox{\scriptsize LI}_{2}} \sim \frac{\theta_{\mbox{\scriptsize E}}}{ (f_{\mbox{\scriptsize L}} / f_{\mbox{\scriptsize S}}) \, u^{5} }
\label{eqn:centroidLI2largeu}
\end{equation}

Up to this point, only the dependence of the microlensing signals
on the normalised source-lens separation $u$ has been discussed. However, $u$ is fundamentally a function of time as the source and lens
move on the celestial sphere relative to each other, principally due to their proper motions and annual parallaxes
(Figure~\ref{fig:microlensing}). The relative motion is typically dominated by the relative proper motion, which leads to
characteristic variations in $u$, and therefore also characteristic variations in the observed
magnification and centroid shift of the microlensing event as a function of time.
For uniform rectilinear (straight line) relative proper motion, and no other relative motion components, the time-dependent
magnification and centroid shift variations are symmetric around the time of closest approach between the source and the lens
(i.e. the time $t = t_{0}$ at which $u = u_{0}$ is a minimum; see Figure~2 in \citealt{pac1986} and Figure~1 in D00).

Astrometric microlensing events are characterised by a slowing of the (apparent) relative source-lens motion as $u$ decreases
towards $u_{0}$, a rapid acceleration and deceleration of the relative motion as $u$ passes through its minimum
value $u = u_{0}$ (including a rotational component to the relative motion), a further slowing of the relative motion as $u$
increases away from $u_{0}$, and a final recovery to the original relative source-lens motion in the absence of lensing
(\citealt{chi1989}; D00). This motion signature can be identified as being due to microlensing, and
correctly distinguished from induced motion by a nearby source companion, as long as the closest approach $u = u_{0}$
occurs during the observational time baseline (\citealt{bel2002}). Even so, the presence of a massive
companion to either the source and/or the lens will make the modelling of an astrometric microlensing event considerably more challenging
(e.g. \citealt{an2002}; \citealt{saj2014}; \citealt{nuc2016}).

\subsection{Examples Specific To \textit{Gaia}}
\label{sec:examplegaia}

\begin{figure*}
\centering
\epsfig{file=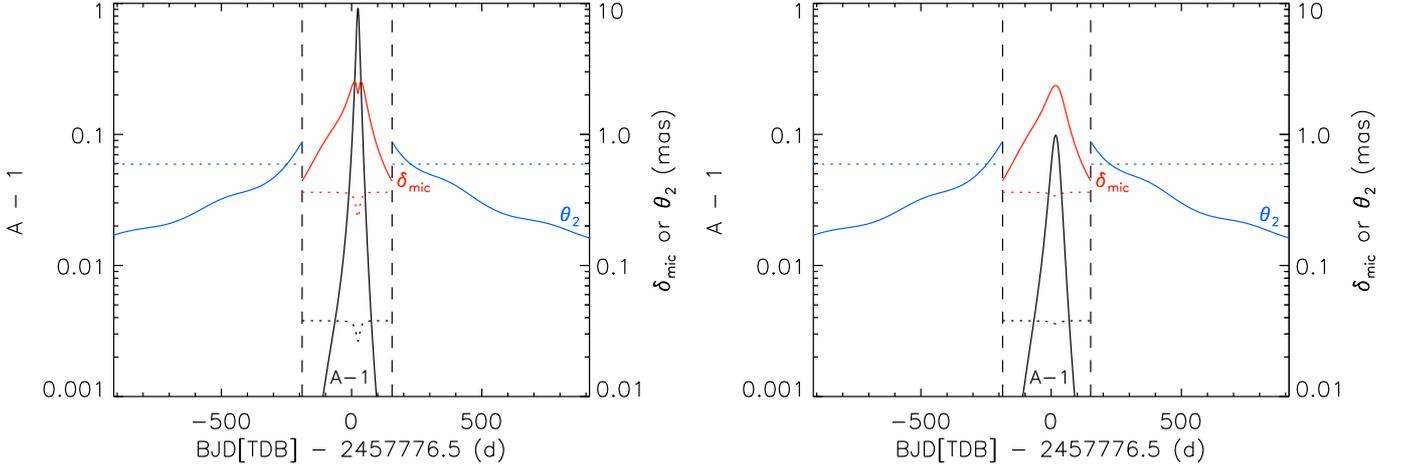,angle=0.0,width=\linewidth}
\caption{Variation of the magnification above baseline $A - 1$ (black curve) and the centroid shifts $\delta_{\mbox{\scriptsize mic}}$
         (mas; red curve) and $\theta_{2}$ (mas; blue curve) as a function of time (d) for the two fictitious microlensing events described in
         Section~\ref{sec:examplegaia}. The events differ only in that $u_{0} = 0.6$ and $u_{0} = 1.6$ for the left-
         and right-hand panels, respectively. The vertical dashed lines in each panel indicate when an event switches between unresolved and
         partially-resolved microlensing based on the median \textit{Gaia} spatial resolution of $\sim$103~mas. The dotted curves
         indicate the photometric ($G$-band) and astrometric (along-scan) precisions for a single observation with black, red and blue
         colours corresponding to $A - 1$, $\delta_{\mbox{\scriptsize mic}}$ and $\theta_{2}$, respectively.
         \label{fig:example}}
\end{figure*}

For microlensing events with $u_{0}>8.5$, the peak values of $A-1$ and $A_{1}-1$
are less than the equivalent of 0.4~mmag above baseline, which is below the
$G$-band photometric precision of \textit{Gaia} for bright stars (Equation~\ref{eqn:photprec}). Even though these signals
cannot be detected in \textit{Gaia} photometry, it is still possible that the associated centroid shifts
$\delta_{\mbox{\scriptsize mic}}$ and $\theta_{2}$ are detectable because, for large and increasing $u$,
they follow a much slower $u^{-1}$ asymptotic decline than the $u^{-4}$ declines of $A-1$ and $A_{1}-1$
(Section~\ref{sec:microlensingbehaviour}). The predicted microlensing event from M18 falls into this category.
The predicted event has $u_{0}\approx11.6$, $\theta_{\mbox{\scriptsize E}}\approx32.8$~mas, and
$\theta_{\mbox{\scriptsize E}} \, u_{0} \approx 380$~mas for a $G\approx18$~mag source star. Since the median spatial resolution of
\textit{Gaia} is $\sim$103~mas (Section~\ref{sec:gaia}), the event will unfold completely in the partially-resolved
microlensing regime. The maximum centroid shift $\theta_{2}\approx2.8$~mas of the major source image I$_{1}$
can be detected by \textit{Gaia}
in a single observation at $\sim$3.6$\,\sigma_{\mbox{\scriptsize AL}}$ for the most favourable scanning angles (Equation~\ref{eqn:modelastromal}),
while the remaining signals $A_{1}-1$, $A_{\mbox{\scriptsize LI}_{2}}-1$ and
$\theta_{\mbox{\scriptsize LI}_{2}}$ are all undetected.
Interestingly, for similar event parameters $u_{0}=8.5$, $\theta_{\mbox{\scriptsize E}}=32.8$~mas,
$\theta_{\mbox{\scriptsize E}} \, u_{0} \approx 279$~mas, and ``swapped''
source and lens $G$ magnitudes of 10 and 18~mag, respectively, the peak signals are
$A_{1}-1\equiv0.2$~mmag ($\sim$0.5$\,\sigma_{G}$; Equation~\ref{eqn:photprec}),
$\theta_{2}\approx3.8$~mas ($\sim$116$\,\sigma_{\mbox{\scriptsize AL}}$),
$A_{\mbox{\scriptsize LI}_{2}}-1\equiv0.27$~mag ($\sim$42$\,\sigma_{G}$), and
$\theta_{\mbox{\scriptsize LI}_{2}} \approx 0.85$~mas ($\sim$1.3$\,\sigma_{\mbox{\scriptsize AL}}$).
In other words, for I$_{1}$, the peak centroid shift is securely detected while the peak magnification above baseline is undetected,
and for the LI$_{2}$ blend, the peak magnification above baseline and centroid shift are securely and border-line detected, respectively.

Taking this last example further, for a dark lens, at $u = u_{0}$, the major and minor source images have a separation of $\sim$286~mas
and they are resolved by \textit{Gaia}. More importantly though, the \textit{minor image is also bright enough to be detected by Gaia} as it
has a $G$ magnitude of $\sim$19.4~mag. Since \textit{Gaia} performs on-the-fly object detection, I strongly recommend
that whenever a microlensing event of a bright source due to an invisible massive object is identified
by \textit{Gaia} (e.g. \citealt{har2018}), the data are checked for a transient object detection near the expected position
of the minor source image. If the minor source image is successfully recorded and measured, then this will be the first time that
a fully resolved lensing event is observed for a source and lens within our Galaxy.

In Figure~\ref{fig:example}, I plot $A - 1$ (black curves), $\delta_{\mbox{\scriptsize mic}}$ (red curves)
and $\theta_{2}$ (blue curves) as a function of time for two fictitious microlensing events.
For both events, I assume that the source star is at a distance of $D_{\mbox{\scriptsize S}} = 4$~kpc.
The lens is taken to be the 0.675~$M_{\sun}$ white dwarf Stein~2051~B (\citealt{sah2017}) moved out to a
distance of $D_{\mbox{\scriptsize L}} = 60$~pc. At this distance, the lens would have an apparent $V$ magnitude
of $\sim$17.6~mag and a proper motion of $\sim$0.218\,\arcsec~yr$^{-1}$.
I assume a source star proper motion of 1~mas~yr$^{-1}$ in the same direction as the lens. I also assume
that the source has the same apparent magnitude as the lens so that
$f_{\mbox{\scriptsize L}} / f_{\mbox{\scriptsize S}} = 1$ (which is the case for a Sun-like star at 4~kpc).
The Einstein radius for this microlensing configuration is $\theta_{\mbox{\scriptsize E}} \approx 9.5$~mas.
In the absence of the parallactic motions, I set $u_{0} = 0.6$ (left-hand panel) and $u_{0} = 1.6$ (right-hand panel),
and this is set to occur at time $t_{0} = 2457776.5$~d (BJD[TDB]; mid-mission for \textit{Gaia}).
Annual parallax is then included in the source and lens motions
by adopting the celestial coordinates of Stein~2051~B and by assuming that the proper motions
are at constant declination (Section~\ref{sec:starpaths}).

For \textit{Gaia}, both the unresolved and partially-resolved microlensing regimes are relevant to these example events.
The vertical dashed lines in each panel of Figure~\ref{fig:example} indicate a regime change.
In the unresolved regime, the blend has a baseline magnitude of $\sim$16.8~mag. The corresponding
photometric ($G$-band) and astrometric (along-scan) precisions for a single observation
(Equations~\ref{eqn:photprec}~and~\ref{eqn:modelastromal})
are plotted in the panels of Figure~\ref{fig:example} as horizontal curves (black, red, and blue for $A-1$,
$\delta_{\mbox{\scriptsize mic}}$, and $\theta_{2}$, respectively).
Both events have essentially the same detectability characteristics despite the different values of $u_{0}$
and the different amplitudes/shapes of the signals at their peaks.
From a single observation,
\textit{Gaia} can recover $A-1$ and $\delta_{\mbox{\scriptsize mic}}$ above 3-sigma
in a time-window of $\sim$100~d centred on $t=t_{0}$. At least two observations are guaranteed during this window.
Hence, for the most favourable scanning angles, the photometric and astrometric signals are securely detected in the unresolved
microlensing regime. In the partially-resolved microlensing regime, $A_{1}-1$, $A_{\mbox{\scriptsize LI}_{2}}-1$
and $\theta_{\mbox{\scriptsize LI}_{2}}$ are negligible,
and these signals go undetected by \textit{Gaia}. However, $\theta_{2}$ is in the range
$\sim$1-1.5$\,\sigma_{\mbox{\scriptsize AL}}$ for time periods of $\sim$65~d immediately before and after the
unresolved regime. Hence one of the astrometric signals has a small chance of being border-line detected in the
partially-resolved regime.

\section{Needles In A Haystack}
\label{sec:methods}

In this section, I describe the methods I use to identify source-lens pairs from GDR2 that could potentially
lead to microlensing events that are detectable by \textit{Gaia} or other observing facilities.
To perform such an analysis, one needs certain ingredients.
The previous section already dealt with microlensing effects as a function of $M$,
$D_{\mbox{\scriptsize L}}$, $D_{\mbox{\scriptsize S}}$, $u$ and $f_{\mbox{\scriptsize L}} / f_{\mbox{\scriptsize S}}$
(or alternatively $\theta_{\mbox{\scriptsize E}}$, $u$ and $f_{\mbox{\scriptsize L}} / f_{\mbox{\scriptsize S}}$).
Further essential ingredients are detailed in Section~\ref{sec:otheringredients},
while the analysis of GDR2 is described in Section~\ref{sec:gdr2analysis}.

\subsection{Ingredients}
\label{sec:otheringredients}

\subsubsection{Stellar Paths Across The Celestial Sphere}
\label{sec:starpaths}

Let the angular position vectors $\vec{\phi}_{\mbox{\scriptsize S}}$ and $\vec{\phi}_{\mbox{\scriptsize L}}$ on the celestial
sphere for the source and lens, respectively, be formulated as functions of time $t$. Taking into account proper motion
and annual parallax, one may write:
\begin{equation}
\begin{aligned}
\vec{\phi}_{\mbox{\scriptsize S}}(t) & = \left( \begin{matrix} \alpha_{\mbox{\scriptsize S}}(t) \\ \delta_{\mbox{\scriptsize S}}(t) \end{matrix} \right) \\
                                     & \approx \left( \begin{matrix} \alpha_{\mbox{\scriptsize ref,S}} \\ \delta_{\mbox{\scriptsize ref,S}} \end{matrix} \right)
                                     + \left( \begin{matrix} \mu_{\alpha *,\mbox{\scriptsize S}}/\cos(\delta_{\mbox{\scriptsize ref,S}}) \\
                                                             \mu_{\delta,\mbox{\scriptsize S}} \end{matrix} \right)
                                       \left[ t - t_{\mbox{\scriptsize ref}} \right] \\
                                     & \;\;\;\; + \varpi_{\mbox{\scriptsize S}}
                                       \left( \begin{matrix} P_{\alpha,\mbox{\scriptsize S}}(t)/\cos(\delta_{\mbox{\scriptsize ref,S}})
                                       \\ P_{\delta,\mbox{\scriptsize S}}(t) \end{matrix} \right) \\
\label{eqn:sourcemotion}
\end{aligned}
\end{equation}
\begin{equation}
\begin{aligned}
\vec{\phi}_{\mbox{\scriptsize L}}(t) & = \left( \begin{matrix} \alpha_{\mbox{\scriptsize L}}(t) \\ \delta_{\mbox{\scriptsize L}}(t) \end{matrix} \right) \\
                                     & \approx \left( \begin{matrix} \alpha_{\mbox{\scriptsize ref,L}} \\ \delta_{\mbox{\scriptsize ref,L}} \end{matrix} \right)
                                     + \left( \begin{matrix} \mu_{\alpha *,\mbox{\scriptsize L}}/\cos(\delta_{\mbox{\scriptsize ref,L}}) \\
                                                             \mu_{\delta,\mbox{\scriptsize L}} \end{matrix} \right)
                                       \left[ t - t_{\mbox{\scriptsize ref}} \right] \\
                                     & \;\;\;\; + \varpi_{\mbox{\scriptsize L}}
                                       \left( \begin{matrix} P_{\alpha,\mbox{\scriptsize L}}(t)/\cos(\delta_{\mbox{\scriptsize ref,L}})
                                       \\ P_{\delta,\mbox{\scriptsize L}}(t) \end{matrix} \right) \\
\label{eqn:lensmotion}
\end{aligned}
\end{equation}
where the subscripts S and L correspond to the source and the lens, respectively. The vector $\vec{\phi}(t)$ has
components of right ascension $\alpha(t)$ and declination $\delta(t)$. The coordinates
$( \alpha_{\mbox{\scriptsize ref}}, \delta_{\mbox{\scriptsize ref}} )$ are celestial coordinates at the reference
epoch $t = t_{\mbox{\scriptsize ref}}$. For GDR2, $t_{\mbox{\scriptsize ref}}=\mbox{J}2015.5=2457206.375$~d (BJD[TDB]).
The quantities $\mu_{\alpha *}$ and $\mu_{\delta}$ are the tangent plane projections
of the proper motion vector in the directions of increasing right ascension and declination, respectively, 
and $\varpi$ is the annual parallax. The functions
$P_{\alpha}(t)$ and $P_{\delta}(t)$ are the parallax factors specified by:
\begin{equation}
P_{\alpha}(t) \approx X(t) \sin(\alpha_{\mbox{\scriptsize ref}}) - Y(t) \cos(\alpha_{\mbox{\scriptsize ref}})
\label{eqn:ra_par_fac}
\end{equation}
\begin{equation}
\begin{aligned}
P_{\delta}(t) \approx & \, X(t) \cos(\alpha_{\mbox{\scriptsize ref}}) \sin(\delta_{\mbox{\scriptsize ref}})
                        + Y(t) \sin(\alpha_{\mbox{\scriptsize ref}}) \sin(\delta_{\mbox{\scriptsize ref}}) \\
                      & - Z(t) \cos(\delta_{\mbox{\scriptsize ref}}) \\
\label{eqn:dec_par_fac}
\end{aligned}
\end{equation}
Here $X(t)$, $Y(t)$ and $Z(t)$ are the Solar-system barycentric coordinates of the Earth in astronomical units at the J2000.0
reference epoch (see Section~7.2.2.3 in \citealt{urb2013}). I used the Jet Propulsion Laboratory HORIZONS on-line ephemeris computation
service\footnote{\url{https://ssd.jpl.nasa.gov/horizons.cgi}} to obtain tabulated values of $X(t)$, $Y(t)$ and $Z(t)$ at daily intervals
over the date ranges relevant to this paper. I then employ cubic spline interpolation to calculate $X(t)$, $Y(t)$ and $Z(t)$ 
from these data for any time $t$.

\subsubsection{Observational Time Baseline Of \textit{Gaia}}
\label{sec:times}

All times $t$ listed in this paper are Barycentric Julian Dates (BJD) in Barycentric Dynamical Time (TDB).
Science observations for \textit{Gaia} started on 25th July 2014 ($t=2456863.5$~d). For the purpose
of predicting microlensing events, I adopt an optimistic end-date for the \textit{Gaia} mission of 25th July 2026 ($t=2461246.5$~d;
maximal mission duration of 12 years).

\subsubsection{Photometric Precision Of \textit{Gaia}}
\label{sec:photprec}

An estimate of the standard error $\sigma_{G}$ on the $G$ magnitude for a single observation, which
includes a 20\% contingency margin for unknown systematic errors, is given by P16 as:
\begin{equation}
\sigma_{G} = 0.0012 \, \sqrt{ 0.04895 \, z^{2} + 1.8633 \, z + 0.0001985 } \,\,\,\, \text{mag}
\label{eqn:photprec}
\end{equation}
where:
\begin{equation}
z = \max \left\{ 10^{0.4 \, (12 - 15)}, 10^{0.4 \, (G - 15)} \right\}
\label{eqn:zphot}
\end{equation}
For stars brighter than $G=12$~mag, these formulae yield $\sigma_{G} \approx 0.4$~mmag.

\subsubsection{Astrometric Precision Of \textit{Gaia}}
\label{sec:astroprec}

\begin{figure*}
\centering
\epsfig{file=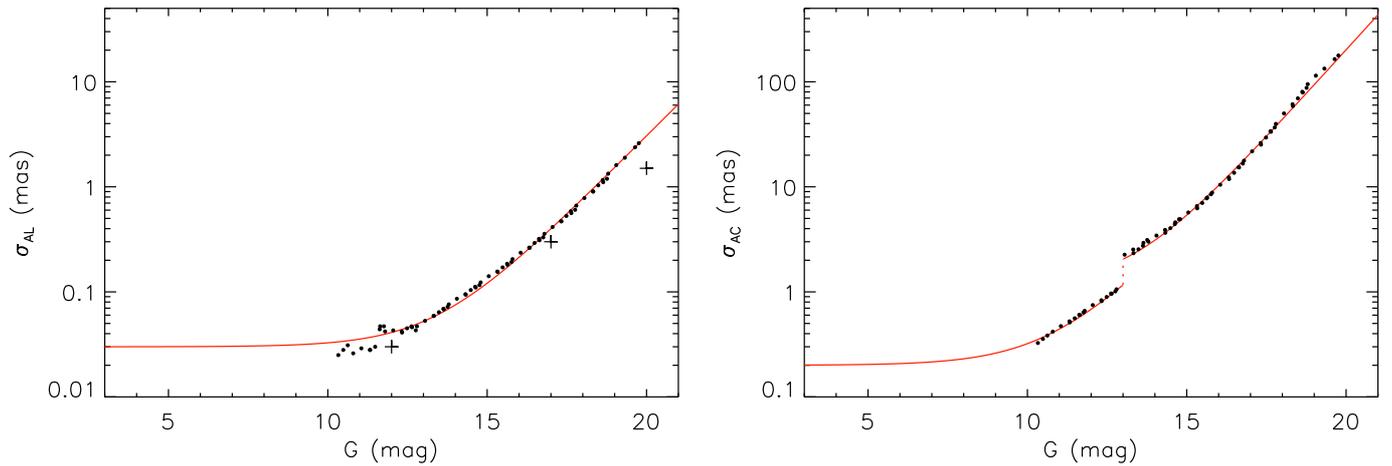,angle=0.0,width=\linewidth}
\caption{Plot of $\sigma_{\mbox{\scriptsize AL}}$ (mas; left-hand panel) and $\sigma_{\mbox{\scriptsize AC}}$
         (mas; right-hand panel) versus $G$ magnitude. The data from R18 are plotted as black dots and the formal
         errors reported by \citet{fab2016} are plotted as plus signs in the left-hand panel. The best fit curves are 
         plotted in red.
         \label{fig:astroprec}}
\end{figure*}

The astrometric precision of \textit{Gaia} is typically quantified by the end-of-mission parallax standard error
(e.g. P16). However, for this work, I instead require an estimate of the astrometric precision for a
single observation. Let the standard errors on the positional measurement for a single observation in the along-
and across-scan directions be denoted by $\sigma_{\mbox{\scriptsize AL}}$ and $\sigma_{\mbox{\scriptsize AC}}$, respectively.
Rybicki et al. (2018; hereafter R18) performed Monte Carlo centroiding simulations to estimate
$\sigma_{\mbox{\scriptsize AL}}$ and $\sigma_{\mbox{\scriptsize AC}}$ as a function
of $V$ magnitude and $V-I$ colour. They note that the colour dependence of their results disappears when considering
$G$ magnitudes. Using the relation between $G-V$ and $V-I$ detailed in Appendix~A of \citet{eva2018}, one may convert
the $V$ and $V-I$ values in Tables~1~and~2 from R18 into $G$ magnitudes. In Figure~\ref{fig:astroprec},
I plot $\sigma_{\mbox{\scriptsize AL}}$ (mas; left-hand panel) and $\sigma_{\mbox{\scriptsize AC}}$ (mas; right-hand panel)
from these tables against $G$ (black dots).

\citet{fab2016} state that the formal error along-scan
for a single astrometric field CCD is 0.06~mas for $G<12$ mag, reaching 0.6~mas at $G=17$~mag, and 3~mas at $G=20$~mag. To compare this
with the results from R18, one must inflate these errors by 50\% and then divide by $\sqrt{9}$ to
account for the combination of independent measurements from nine CCDs. These adjusted errors are plotted in the
left-hand panel of Figure~\ref{fig:astroprec} (plus signs)
where it is clear that they are somewhat smaller than the values from R18. Erring on the side of caution,
I choose to fit the data from R18 and I employ functions of the form $f(G) = \sigma_{\mbox{\scriptsize min}} + B \, \exp \left[ C \, (G - 12) \right]$,
where $\sigma_{\mbox{\scriptsize min}}$, $B$ and $C$ are parameters to be determined.
The data for $\sigma_{\mbox{\scriptsize AC}}$ exhibit a jump\footnote{The jump is caused by the fact that the majority
of objects fainter than $G=13$~mag are binned in the across-scan direction on the nine astrometric field CCDs, which leaves
only the sky mapper CCD to provide the across-scan astrometric measurement for these objects.}
at $G\approx13$~mag, and accordingly I fit the data on either side of the jump independently.
For $\sigma_{\mbox{\scriptsize AL}}$, $\sigma_{\mbox{\scriptsize AC}}$ with $G\le13$~mag, and $\sigma_{\mbox{\scriptsize AC}}$ with $G>13$~mag,
the noise-floor parameter $\sigma_{\mbox{\scriptsize min}}$ is fixed at 0.030, 0.200, and 1.140~mas, respectively.
The fits are plotted in Figure~\ref{fig:astroprec} using red curves.
The corresponding formulae for estimating $\sigma_{\mbox{\scriptsize AL}}$ and
$\sigma_{\mbox{\scriptsize AC}}$ as a function of $G$ magnitude are:
\begin{equation}
\sigma_{\mbox{\scriptsize AL}} = 0.030 + 0.0111 \, \exp \left[ 0.701 \, (G - 12) \right] \,\,\,\, \text{mas}
\label{eqn:modelastromal}
\end{equation}
\begin{equation}
\sigma_{\mbox{\scriptsize AC}} =
\begin{cases}
0.200 + 0.483 \, \exp \left[ 0.690 \, (G - 12) \right] \,\,\,\, \text{mas} & \mbox{for $G\le13$} \\
1.140 + 0.420 \, \exp \left[ 0.771 \, (G - 12) \right] \,\,\,\, \text{mas} & \mbox{for $G>13$}   \\
\end{cases}
\label{eqn:modelastromac}
\end{equation}

\subsection{Selecting Source-Lens Pairs}
\label{sec:gdr2analysis}

Throughout this section and the rest of the paper, GDR2 data column names are written in {\tt TYPEWRITER} font.

\subsubsection{Corrections To GDR2 Quantities}
\label{sec:gdr2}

The astrometric solution for GDR2 is described in \citet{lin2018}, and they found that there is an overall
negative bias in the parallax measurements of $-$0.029~mas. Furthermore, the published uncertainties on the
parameters of the astrometric solutions are underestimated by different amounts for different object populations,
ranging anywhere from $\sim$7 to 50\% (\citealt{bro2018}). Consequently, I correct the GDR2 parallaxes by adding 0.029~mas
to the {\tt PARALLAX} entries, and I inflate the uncertainties on the astrometric parameters by 25\%
(i.e. the entries {\tt RA\_ERROR}, {\tt DEC\_ERROR}, {\tt PMRA\_ERROR}, {\tt PMDEC\_ERROR} and {\tt PARALLAX\_ERROR}).
The astrometric parameter covariances are provided in GDR2 as correlation coefficients, and therefore no
adjustment of these entries is required. The inflated uncertainties are conservative for
faint objects ($G>16$~mag) outside the Galactic plane, and also for objects brighter than $G=12$~mag.
However, for $12<G<15$~mag, the uncertainties may still be too small for some objects. Hereafter,
references to \textit{Gaia} parallax measurements and the astrometric parameter uncertainties
always refer to the corrected parallax values and the inflated uncertainties.

\subsubsection{Lens Star Selection}
\label{sec:lensstars}

\begin{table*}
\centering
\caption{List of constraints that need to be satisfied in order for a GDR2 object to be selected as a lens star.
         The symbols $\varpi$ and $\sigma[\varpi]$ represent the corrected parallax and its inflated uncertainty, respectively.
         Furthermore, $\chi^{2} = {\tt ASTROMETRIC\_CHI2\_AL}$ and $\nu = {\tt ASTROMETRIC\_N\_GOOD\_OBS\_AL} - 5$.}
\small{
\begin{tabular}{@{}lcccl}
\hline
GDR2 Column Name                      & Relation & Value  & Unit & Description \\
\hline
{\tt DUPLICATED\_SOURCE}              & =        & FALSE  & -    & Reject duplicated objects that are likely to have \\
                                      &          &        &      & astrometric and photometric problems \\
{\tt FRAME\_ROTATOR\_OBJECT\_TYPE}    & =        & 0      & -    & Reject known extra-galactic objects \\
{\tt ASTROMETRIC\_PARAMS\_SOLVED}     & =        & 31     & -    & Only accept objects that have a 5-parameter \\
                                      &          &        &      & astrometric solution \\
$\varpi / \sigma[\varpi]$             & >        & 4      & -    & Only accept objects with a sufficiently precise parallax \\
                                      &          &        &      & measurement \\
$\varpi$                              & >        & 0      & mas  & Reject objects with a non-positive parallax \\
$\varpi$                              & <        & 769    & mas  & Reject objects with a parallax greater than that of \\
                                      &          &        &      & Proxima Centauri, which has a \textit{Gaia} parallax \\
                                      &          &        &      & of 768.529$\pm$0.254~mas \\
$\sqrt{\chi^{2} / \nu}$               & <        & $1.2 \, \max \left\{ 1, \exp \left[ 0.2 \, (19.5 - G) \right] \right\}$ & - &
                                                                   Reject objects with a spurious astrometric solution \\
{\tt ASTROMETRIC\_EXCESS\_NOISE\_SIG} & <        & 3      & -    & Reject objects with significant excess noise in the \\
                                      &          &        &      & astrometric solution \\
\hline
\end{tabular}
}
\label{tab:lens_req}
\end{table*}

The first task in identifying potential microlensing events from GDR2 is to select a set of
lens stars from the data. For a reliable microlensing prediction, a lens must have a well-constrained mass,
distance (or equivalently, parallax), and path on the sky. To meet these requirements,
I conservatively select lens stars from GDR2 using the constraints listed in Table~\ref{tab:lens_req}.
The constraints on {\tt ASTROMETRIC\_PARAMS\_SOLVED} and the parallax signal-to-noise (S/N) ratio $\varpi / \sigma[\varpi]$
limit the lens sample to those stars with well-determined distances and paths on the sky, while the filter on {\tt DUPLICATED\_SOURCE} 
ensures that the astrometric solutions are derived from reliable data. Furthermore, the constraint on $\sqrt{\chi^{2} / \nu}$
has been proposed in Appendix~C of \citet{lin2018}, and adopted by
\citet{are2018} and \citet{bab2018}, as a way of rejecting stars with spurious astrometric solutions.
While inspecting plots of {\tt ASTROMETRIC\_EXCESS\_NOISE\_SIG} versus mean $G$ magnitude ({\tt PHOT\_G\_MEAN\_MAG}),
it also became clear that a filter on this quantity with an appropriate threshold is required as suggested in Chapter~14 of the GDR2 documentation (\citealt{gai2018}).
This constraint rejects lens stars whose motion is not well-described by proper motion and parallax
alone (e.g. binary stars with periods of up to $\sim$10$\times$ the GDR2 time baseline). Microlensing predictions from such stars would be unreliable.
Note that although the constraint on non-positive parallaxes is actually contained within the constraint on the parallax
S/N ratio, it is listed in Table~\ref{tab:lens_req} for added clarity on the implemented filters. Finally, it stands to reason that lens
stars cannot be extra-galactic, which justifies the filter on {\tt FRAME\_ROTATOR\_OBJECT\_TYPE},
and they also cannot be closer to us than our closest neighbour Proxima Centauri.
Application of these constraints to GDR2 yields $N_{\mbox{\scriptsize L}}$=132,944,121 potential lens stars.

\subsubsection{Source Star Selection}
\label{sec:sourcestars}

\begin{table*}
\centering
\caption{List of constraints that need to be satisfied in order for a GDR2 object with a 5-parameter astrometric
         solution to be selected as a source star.
         The symbol $\varpi$ represents the corrected parallax.
         Furthermore, $\chi^{2} = {\tt ASTROMETRIC\_CHI2\_AL}$ and $\nu = {\tt ASTROMETRIC\_N\_GOOD\_OBS\_AL} - 5$.}
\small{
\begin{tabular}{@{}lcccl}
\hline
GDR2 Column Name                      & Relation & Value  & Unit & Description \\
\hline
{\tt DUPLICATED\_SOURCE}              & =        & FALSE  & -    & Reject duplicated objects that are likely to have \\
                                      &          &        &      & astrometric and photometric problems \\
{\tt ASTROMETRIC\_PARAMS\_SOLVED}     & =        & 31     & -    & Only accept objects that have a 5-parameter \\
                                      &          &        &      & astrometric solution \\
$\varpi$                              & <        & 769    & mas  & Reject objects with a parallax greater than that of \\
                                      &          &        &      & Proxima Centauri, which has a \textit{Gaia} parallax \\
                                      &          &        &      & of 768.529$\pm$0.254~mas \\
$\sqrt{\chi^{2} / \nu}$               & <        & $1.2 \, \max \left\{ 1, \exp \left[ 0.2 \, (19.5 - G) \right] \right\}$ & - &
                                                                   Reject objects with a spurious astrometric solution \\
{\tt ASTROMETRIC\_EXCESS\_NOISE\_SIG} & <        & 3      & -    & Reject objects with significant excess noise in the \\
                                      &          &        &      & astrometric solution \\
\hline
\end{tabular}
}
\label{tab:source_req5}
\end{table*}

\begin{table*}
\centering
\caption{List of constraints that need to be satisfied in order for a GDR2 object with a 2-parameter astrometric
         solution to be selected as a source star.
         Here, $\chi^{2} = {\tt ASTROMETRIC\_CHI2\_AL}$ and $\nu = {\tt ASTROMETRIC\_N\_GOOD\_OBS\_AL} - 2$.
         Note that the constraint on $\sqrt{\chi^{2} / \nu}$ is somewhat more relaxed than the similar constraint on
         objects with a 5-parameter astrometric solution (see Tables~\ref{tab:lens_req}~and~\ref{tab:source_req5}).}
\small{
\begin{tabular}{@{}lcccl}
\hline
GDR2 Column Name                      & Relation & Value  & Unit & Description \\
\hline
{\tt DUPLICATED\_SOURCE}              & =        & FALSE  & -    & Reject duplicated objects that are likely to have \\
                                      &          &        &      & astrometric and photometric problems \\
{\tt ASTROMETRIC\_PARAMS\_SOLVED}     & =        & 3      & -    & Only accept objects that have a 2-parameter \\
                                      &          &        &      & astrometric solution \\
{\tt VISIBILITY\_PERIODS\_USED}       & $\ge$    & 5      & -    & Reject objects with too few independent groups of \\
                                      &          &        &      & observations (at least 4 days between each group) \\
$\sqrt{\chi^{2} / \nu}$               & <        & $1.2 \, 
\max \left\{ 1.5, \exp \left[ 0.2 \, (19.5 - G) \right] \right\}$ & - &
                                                                   Reject objects with a spurious astrometric solution \\
{\tt ASTROMETRIC\_EXCESS\_NOISE\_SIG} & <        & 4      & -    & Reject objects with significant excess noise in the \\
                                      &          &        &      & astrometric solution \\
\hline
\end{tabular}
}

\label{tab:source_req2}
\end{table*}

I also use GDR2 as a catalogue of source stars. The constraints that I adopt for selecting source stars
are somewhat more relaxed than those that I adopt for lens stars. This is mainly because source stars are likely
to be much more distant than lens stars for the majority of microlensing geometries, and the parallax signal
for the source star in such cases may be too small for \textit{Gaia} to detect with a reasonable S/N ratio.
Fortunately, when $\varpi_{\mbox{\scriptsize S}} \ll \varpi_{\mbox{\scriptsize L}}$,
the estimated size of the Einstein ring is not affected by a lack of the
source parallax measurement and corresponding distance estimate (Equation~\ref{eqn:einsteinradius} and noting that $\varpi=1/D$).
Therefore I do not place any constraints on whether an object has a 2- or 5-parameter astrometric solution,
or on the parallax S/N ratio $\varpi / \sigma[\varpi]$, when selecting potential source stars.

For stars with a 5-parameter astrometric solution, I select source stars from GDR2 using the
constraints listed in Table~\ref{tab:source_req5}. Compared to the constraints listed in Table~\ref{tab:lens_req} for lens stars,
the filter on {\tt FRAME\_ROTATOR\_OBJECT\_TYPE} has been dropped, since extra-galactic sources can be lensed,
and the constraint on $\varpi / \sigma[\varpi]$ has also been dropped 
for the reasons already stated. Application of these constraints to GDR2 yields 1,127,324,197 potential source stars.

For stars with only a 2-parameter astrometric solution, I select source stars from GDR2 using the constraints
listed in Table~\ref{tab:source_req2}. Most stars with only a 2-parameter astrometric solution are faint
($G>20$~mag) and I found it necessary to relax the thresholds on $\sqrt{\chi^{2} / \nu}$ and
${\tt ASTROMETRIC\_EXCESS\_NOISE\_SIG}$ by inspecting the distributions of these quantities as a function of
mean $G$ magnitude. I also introduce a filter on {\tt VISIBILITY\_PERIODS\_USED} similar to that which
is applied by \citet{lin2018} for accepting 5-parameter astrometric solutions.
Application of these constraints to GDR2 yields 238,748,126 more source stars,
for a grand total of $N_{\mbox{\scriptsize S}}$=1,366,072,323 potential source stars.

\subsubsection{Initial Source-Lens Pair Selection}
\label{sec:initsourcelenspairs}

Finding the microlensing events that will occur during the mission lifetime of \textit{Gaia} from GDR2 is essentially
a very large cross-matching problem between the $N_{\mbox{\scriptsize S}}$ potential source stars and $N_{\mbox{\scriptsize L}}$ potential lens
stars already selected from the catalogue. Its solution requires considering
$N_{\mbox{\scriptsize L}} (N_{\mbox{\scriptsize S}} - 1)$ source-lens pairs, which is an extremely large number.
Fortunately, one can immediately use spatial information to reject the vast majority of pairs.
A conservative upper limit $\theta_{\mbox{\scriptsize E,max}}$ on the value of the Einstein radius for any particular source-lens pair can be
estimated by assuming a maximum lens mass of $M_{\mbox{\scriptsize max}}=10M_{\sun}$, a maximum lens parallax of
$\varpi_{\mbox{\scriptsize L,max}}=\varpi_{\mbox{\scriptsize L}}+3\sigma[\varpi_{\mbox{\scriptsize L}}]$
(where $\sigma[\varpi_{\mbox{\scriptsize L}}]$ is the uncertainty on $\varpi_{\mbox{\scriptsize L}}$),
and a source parallax of zero (Equation~\ref{eqn:einsteinradius}).
To translate $\theta_{\mbox{\scriptsize E,max}}$ into a maximum source-lens angular separation $\theta_{\mbox{\scriptsize det}}$
within which a microlensing signal can be detected, I only consider the astrometric signals, since, for large and
increasing $u$, they decline at a much slower rate than the photometric signals. Specifically, for large $u$, 
the deflection $\theta_{2}$ of the major source image has the greatest amplitude. The best astrometric precision
achievable by any current observing facility in a single observation (excluding radio interferometry)
is \textit{Gaia}'s $\sim$0.030~mas precision
in the along-scan direction for bright objects (Equation~\ref{eqn:modelastromal}).
Requiring $\theta_{2} \ge 0.030$~mas, and using Equations~\ref{eqn:theta1}~and~\ref{eqn:theta2}, one obtains an upper limit
$\theta_{\mbox{\scriptsize det}} = \theta_{\mbox{\scriptsize E,max}}^{\,2} / 0.030$~mas.

For each potential lens star, I calculate the value of $\theta_{\mbox{\scriptsize det}}$.
The minimum, median, and maximum values
of $\theta_{\mbox{\scriptsize det}}$ over all lens stars are $\sim$0.27, 2.75, and 2092~arcsec, respectively.
To account (very) conservatively for source and lens motions, and for errors in the astrometric parameters,
I compute the following quantity for each source-lens pair:
\begin{equation}
\begin{aligned}
\theta^{\prime}_{\mbox{\scriptsize det}} = & \,\, \theta_{\mbox{\scriptsize det}}
                                             + 3\sigma[\alpha_{*,\mbox{\scriptsize ref,S}}] + 3\sigma[\alpha_{*,\mbox{\scriptsize ref,L}}]
                                             + 3\sigma[\delta_{\mbox{\scriptsize ref,S}}] + 3\sigma[\delta_{\mbox{\scriptsize ref,L}}] \\
                                           & + T_{\mbox{\scriptsize rem}} \, \left( \,|\,\mu_{\alpha*,\mbox{\scriptsize S}}| + 3\sigma[\mu_{\alpha*,\mbox{\scriptsize S}}]
                                                                          + |\,\mu_{\alpha*,\mbox{\scriptsize L}}| + 3\sigma[\mu_{\alpha*,\mbox{\scriptsize L}}] \,\right) \\
                                           & + T_{\mbox{\scriptsize rem}} \, \left( \, |\,\mu_{\delta,\mbox{\scriptsize S}}| + 3\sigma[\mu_{\delta,\mbox{\scriptsize S}}]   
                                                                          + |\,\mu_{\delta,\mbox{\scriptsize L}}| + 3\sigma[\mu_{\delta,\mbox{\scriptsize L}}] \,\right)   \\
                                           & + \varpi_{\mbox{\scriptsize S}} + 3\sigma[\varpi_{\mbox{\scriptsize S}}]
                                             + \varpi_{\mbox{\scriptsize L}} + 3\sigma[\varpi_{\mbox{\scriptsize L}}]                                           \\
\label{eqn:newlim}
\end{aligned}
\end{equation}
where $\sigma[\alpha_{*,\mbox{\scriptsize ref}}] / \cos(\delta_{\mbox{\scriptsize ref}})$, $\sigma[\delta_{\mbox{\scriptsize ref}}]$,
$\sigma[\mu_{\alpha*}]$, $\sigma[\mu_{\delta}]$, and $\sigma[\varpi]$
are the uncertainties on $\alpha_{\mbox{\scriptsize ref}}$ ({\tt RA}), $\delta_{\mbox{\scriptsize ref}}$ ({\tt DEC}),
$\mu_{\alpha*}$ ({\tt PMRA}), $\mu_{\delta}$ ({\tt PMDEC}),
and $\varpi$ ({\tt PARALLAX}), respectively, and $T_{\mbox{\scriptsize rem}}\approx11.1$~years is
the maximum remaining time for the \textit{Gaia} mission as counted from the GDR2 reference epoch (Section~\ref{sec:times}).
I reject all source-lens pairs for which the angular distance between them at $t=t_{\mbox{\scriptsize ref}}$
exceeds $\theta^{\prime}_{\mbox{\scriptsize det}}$.
This leaves 85,135,565 source-lens pairs, with 22,616,138 unique lenses, for further consideration, which is a
much more manageable number.

\subsubsection{Refined Source-Lens Pair Selection}
\label{sec:refsourcelenspairs}

The list of source-lens pairs can be filtered further by considering more carefully the upper limit on the lens mass $M_{\mbox{\scriptsize max}}$
on a case-by-case basis.
For a main sequence lens star more massive than the Sun, the relation $L/L_{\sun}\approx(M/M_{\sun})^{4}$ holds.
Using the lens parallax and its mean $G$ magnitude, and assuming an
absolute bolometric magnitude for the Sun of $M_{\mbox{\scriptsize bol},\sun}\approx4.74$~mag, one may estimate the lens
luminosity, and therefore its mass. The fact that extinction and bolometric corrections have been ignored imply
that the mass is underestimated. Accounting for ample extinction ($A_{G}\approx2$~mag) and maximal bolometric corrections ($\mbox{BC}\approx-4$~mag),
I set $M_{\mbox{\scriptsize max}}$ to $10^{0.6}\approx4$~times the lens mass estimate. For giant stars, $M_{\mbox{\scriptsize max}}$
is over-estimated since they are more luminous than main sequence stars, and hence the value of $M_{\mbox{\scriptsize max}}$ computed in this way
also serves as a maximum lens mass for giant stars. If $M_{\mbox{\scriptsize max}}$ is less than the Chandrasekhar limit, then I increase it to
1.44$M_{\sun}$ to cover the possibility that the lens is a white dwarf, which also serves as an upper limit to the lens
mass for sub-solar mass main sequence stars and brown dwarfs. Using this improved upper limit on the lens
mass $M_{\mbox{\scriptsize max}}$, and again adopting $\varpi_{\mbox{\scriptsize L,max}}=\varpi_{\mbox{\scriptsize L}}+3\sigma[\varpi_{\mbox{\scriptsize L}}]$
and a source parallax of zero, I calculate a new value of $\theta_{\mbox{\scriptsize E,max}}$ for each source-lens pair.

Considering the six possible microlensing signals $A$, $\delta_{\mbox{\scriptsize mic}}$,
$A_{1}$, $\theta_{2}$, $A_{\mbox{\scriptsize LI}_{2}}$, and $\theta_{\mbox{\scriptsize LI}_{2}}$, one finds that
their asymptotic behaviours for $u\gg1$, as described by Equations~\ref{eqn:maglargeu}~to~\ref{eqn:centroidLI2largeu}, always bound-above their
corresponding exact expressions in Equations~\ref{eqn:magnification}, \ref{eqn:centroidshift}, \ref{eqn:a1},
\ref{eqn:theta2}, \ref{eqn:ALI2}, and \ref{eqn:centroidLI2}, respectively. By comparing the asymptotic amplitude of each signal to
a best-achievable photometric or astrometric precision, then one may compute
a new maximum source-lens angular separation $\theta_{\mbox{\scriptsize det}}$ within which at least one of the signals can be detected.
The bright-limit photometric precision for \textit{Gaia} of $\sigma_{G}\approx0.4$~mmag for a single observation in the
$G$-band (Equation~\ref{eqn:photprec}) is also widely applicable to ground-based telescopes and
it is of the correct order of magnitude for various space telescopes
(e.g. \textit{HST}). The bright-limit astrometric precision for \textit{Gaia} of
$\sigma_{\mbox{\scriptsize AL}}\approx0.030$~mas for a single observation in the along-scan direction (Equation~\ref{eqn:modelastromal})
is the best for any current observing facility, although it requires a very specific orientation of the scanning angle.
Averaging over all possible scanning angles, the bright-limit astrometric precision per-observation for \textit{Gaia} is $\sim$0.131~mas
(Equations~\ref{eqn:modelastromal}~and~\ref{eqn:modelastromac}),
which is similar to the best astrometric precision achievable by \textit{HST} of $\sim$0.2~mas (e.g. \citealt{kai2017}).
Therefore, I adopt the following requirements for each signal:
\begin{equation}
A , A_{1} , A_{\mbox{\scriptsize LI}_{2}} \ge 1 + \sigma_{\mbox{\scriptsize amp}} = 10^{0.4(0.0004)} \approx 1.00037
\label{eqn:req1}
\end{equation}
\begin{equation}
\delta_{\mbox{\scriptsize mic}} , \theta_{2} , \theta_{\mbox{\scriptsize LI}_{2}} \ge \sigma_{\mbox{\scriptsize ast}} = 0.131 \,\, \text{mas}
\label{eqn:req2}
\end{equation}
By using Equations~\ref{eqn:maglargeu}~to~\ref{eqn:centroidLI2largeu}, the following upper limits can then be derived:
\begin{equation}
\theta_{\mbox{\scriptsize det},A} = 2^{1/4} \, (1 + f_{\mbox{\scriptsize L}} / f_{\mbox{\scriptsize S}})^{-1/4}
                                    \, \sigma_{\mbox{\scriptsize amp}}^{-1/4}
                                    \, \theta_{\mbox{\scriptsize E,max}}
\label{eqn:lim1}
\end{equation}
\begin{equation}
\theta_{\mbox{\scriptsize det},\delta\mbox{\scriptsize mic}} = (1 + f_{\mbox{\scriptsize L}} / f_{\mbox{\scriptsize S}})^{-1}
                                                               \, \sigma_{\mbox{\scriptsize ast}}^{-1}
                                                               \, \theta_{\mbox{\scriptsize E,max}}^{\,2}
\label{eqn:lim2}
\end{equation}
\begin{equation}
\theta_{\mbox{\scriptsize det},A1} = \sigma_{\mbox{\scriptsize amp}}^{-1/4} \, \theta_{\mbox{\scriptsize E,max}}
\label{eqn:lim3}
\end{equation}
\begin{equation}
\theta_{\mbox{\scriptsize det},\theta2} = \sigma_{\mbox{\scriptsize ast}}^{-1} \, \theta_{\mbox{\scriptsize E,max}}^{\,2}
\label{eqn:lim4}
\end{equation}
\begin{equation}
\theta_{\mbox{\scriptsize det},A\mbox{\scriptsize LI}2} =  (f_{\mbox{\scriptsize L}} / f_{\mbox{\scriptsize S}})^{-1/4}
                                                           \, \sigma_{\mbox{\scriptsize amp}}^{-1/4}
                                                           \, \theta_{\mbox{\scriptsize E,max}}
\label{eqn:lim5}
\end{equation}
\begin{equation}
\theta_{\mbox{\scriptsize det},\theta\mbox{\scriptsize LI}2} = (f_{\mbox{\scriptsize L}} / f_{\mbox{\scriptsize S}})^{-1/5}
                                                               \, \sigma_{\mbox{\scriptsize ast}}^{-1/5}
                                                               \, \theta^{\,6/5}_{\mbox{\scriptsize E,max}}
\label{eqn:lim6}
\end{equation}
For $\theta_{\mbox{\scriptsize det}}$, I adopt:
\begin{equation}
\theta_{\mbox{\scriptsize det}} = \max\left\{\theta_{\mbox{\scriptsize det},A},\theta_{\mbox{\scriptsize det},\delta\mbox{\scriptsize mic}},
                                             \theta_{\mbox{\scriptsize det},A1},\theta_{\mbox{\scriptsize det},\theta2},
                                             \theta_{\mbox{\scriptsize det},A\mbox{\scriptsize LI}2},\theta_{\mbox{\scriptsize det},\theta\mbox{\scriptsize LI}2}\right\}
\label{eqn:adopt2}
\end{equation}
The minimum, median, and maximum values of $\theta_{\mbox{\scriptsize det}}$ over all source-lens pairs
are $\sim$0.13, 0.59, and 68.8~arcsec, respectively.

For each source-lens pair, I consider their paths on the sky during the maximal time baseline of the \textit{Gaia} mission (Section~\ref{sec:starpaths}).
I reject all source-lens pairs that do not approach each other to within an angular distance of less than $\theta_{\mbox{\scriptsize det}}$
during this time window. After performing this highly efficient, and yet conservative, filtering step for each source-lens pair,
I have 51,379 source-lens pairs remaining, with 14,067 unique lenses.

\subsubsection{Spurious Astrometric Solutions}
\label{sec:spurious}

\begin{figure}
\centering
\epsfig{file=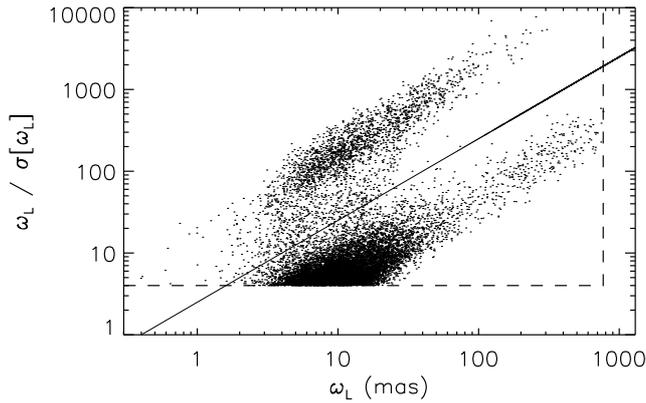,angle=0.0,width=\linewidth}
\caption{Plot of parallax S/N ratio $\varpi_{\mbox{\scriptsize L}}/\sigma[\varpi_{\mbox{\scriptsize L}}]$
         against parallax $\varpi_{\mbox{\scriptsize L}}$ for the lens stars in a set of 51,379 source-lens pairs.
         The continuous line corresponds to $\sigma[\varpi_{\mbox{\scriptsize L}}]=0.4$~mas. Rejected lenses
         fall below this line. The dashed lines representing
         $\varpi_{\mbox{\scriptsize L}}/\sigma[\varpi_{\mbox{\scriptsize L}}]=4$
         and $\varpi_{\mbox{\scriptsize L}}=769$~mas correspond to the constraints applied to GDR2 to select the initial
         lens sample (Table~\ref{tab:lens_req}).
         \label{fig:lensparallaxstn}}
\end{figure}

By exploring scatter plots of various quantities from GDR2 for the lens sample in the latest set of source-lens pairs,
I found that the majority of these lenses have spurious astrometric solutions. This
is most easily observed in Figure~\ref{fig:lensparallaxstn} where the lens parallax S/N ratio
$\varpi_{\mbox{\scriptsize L}}/\sigma[\varpi_{\mbox{\scriptsize L}}]$ is plotted against lens parallax
$\varpi_{\mbox{\scriptsize L}}$.  There is a well-defined group of lenses for which the parallax error is $\sim$30-50 times
larger than that of other lenses at the same parallax. Further investigation of this group of lenses revealed that they
typically lie in highly crowded fields near the Galactic plane or the Magellanic clouds where \textit{Gaia} struggles to
perform reliable astrometric measurements. I therefore clean the sample by rejecting all source-lens pairs
with $\sigma[\varpi_{\mbox{\scriptsize L}}]>0.4$~mas. This limit is plotted in Figure~\ref{fig:lensparallaxstn} as
a continuous straight line, and the cut leaves me with 2,882 source-lens pairs, including 2,600 unique lenses.

\subsubsection{Binary And Co-Moving Stars}
\label{sec:binary}

\begin{figure}
\centering
\epsfig{file=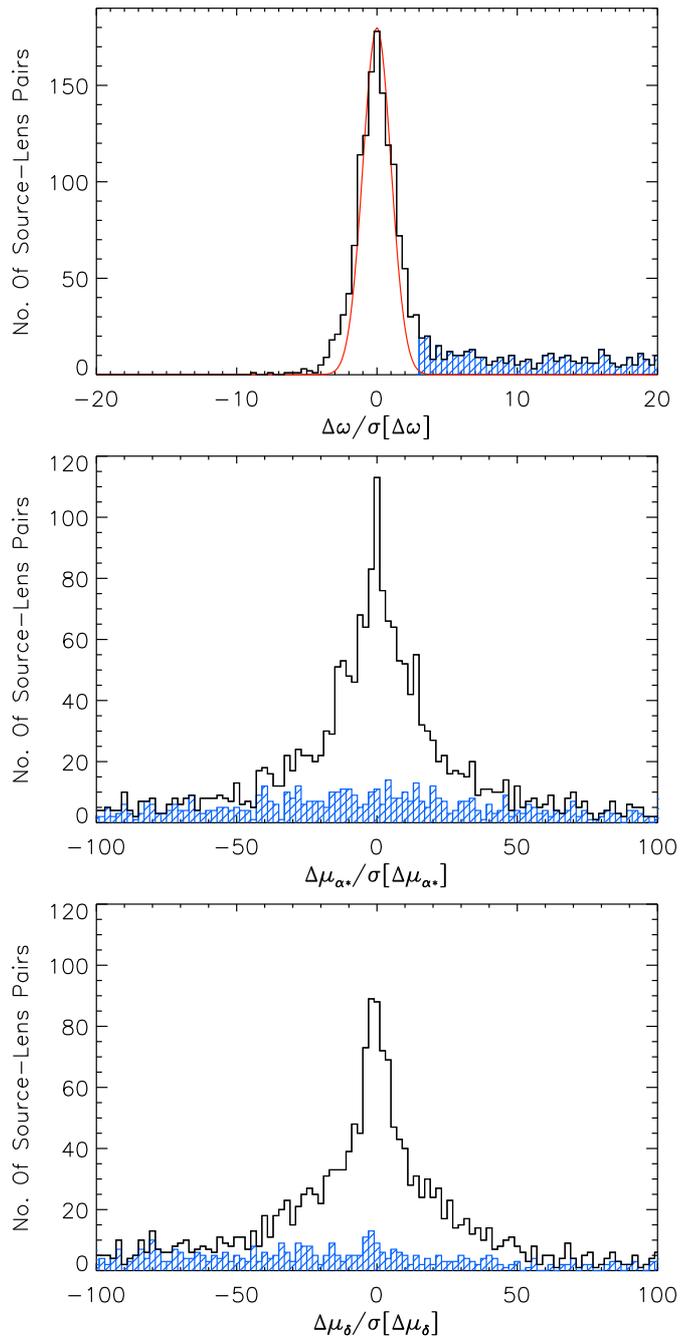,angle=0.0,width=\linewidth}
\caption{Histograms of $\Delta\varpi/\sigma[\Delta\varpi]$ (top panel), $\Delta\mu_{\alpha*}/\sigma[\Delta\mu_{\alpha*}]$ (middle panel)
         and $\Delta\mu_{\delta}/\sigma[\Delta\mu_{\delta}]$ (bottom panel) for 2,882 source-lens pairs (black histograms).
         In the top panel, the red curve is a Gaussian with zero mean and standard deviation of unity, scaled to
         match the histogram of $\Delta\varpi/\sigma[\Delta\varpi]$. In each panel, the blue histogram with striped bars
         is for the source-lens pairs satisfying $\Delta\varpi/\sigma[\Delta\varpi]>3$.
         \label{fig:astrohist}}
\end{figure}

At this stage, it became clear that about half of the source-lens pairs so far selected are either binary stars or co-moving stars at
approximately the same distance from the Sun. Furthermore, the parallaxes indicate that for some of the source-lens pairs,
the lens is more distant than the source. Neither of these source-lens geometries will produce a microlensing event.

In Figure~\ref{fig:astrohist}, I plot histograms of relative source-lens parallax S/N ratio, and relative source-lens proper motion S/N ratio
for each coordinate direction (black histograms). These quantities are defined by:
\begin{equation}
\frac{\Delta\varpi}{\sigma[\Delta\varpi]} = \frac{ \varpi_{\mbox{\scriptsize L}} - \varpi_{\mbox{\scriptsize S}} }
                                                 { (\sigma[\varpi_{\mbox{\scriptsize L}}]^{2} + \sigma[\varpi_{\mbox{\scriptsize S}}]^{2})^{1/2} }
\label{eqn:relparsn}
\end{equation}
\begin{equation}
\frac{\Delta\mu_{\alpha*}}{\sigma[\Delta\mu_{\alpha*}]} = \frac{ \mu_{\alpha*,\mbox{\scriptsize L}} - \mu_{\alpha*,\mbox{\scriptsize S}} }
                                        { (\sigma[\mu_{\alpha*,\mbox{\scriptsize L}}]^{2} + \sigma[\mu_{\alpha*,\mbox{\scriptsize S}}]^{2})^{1/2} }
\label{eqn:relpmrasn}
\end{equation}
\begin{equation}
\frac{\Delta\mu_{\delta}}{\sigma[\Delta\mu_{\delta}]} = \frac{ \mu_{\delta,\mbox{\scriptsize L}} - \mu_{\delta,\mbox{\scriptsize S}} }
                                        { (\sigma[\mu_{\delta,\mbox{\scriptsize L}}]^{2} + \sigma[\mu_{\delta,\mbox{\scriptsize S}}]^{2})^{1/2} }
\label{eqn:relpmdecsn}
\end{equation}
In the top panel, the histogram of $\Delta\varpi/\sigma[\Delta\varpi]$ shows a Gaussian-like core that is a little
wider than a Gaussian with zero mean and standard deviation of unity (red curve). This implies that many of the source-lens
pairs consist of stars at the same distance from the Sun, and that the parallax errors are only slightly under-estimated.
In the middle and bottom panels, the histograms of
$\Delta\mu_{\alpha*}/\sigma[\Delta\mu_{\alpha*}]$ and $\Delta\mu_{\delta}/\sigma[\Delta\mu_{\delta}]$
show the broad peaked distributions expected for a sample of source-lens pairs including binary and co-moving stars.

To filter out binary and co-moving source-lens pairs, I reject all source-lens pairs with
$\Delta\varpi/\sigma[\Delta\varpi]<3$. In each panel
of Figure~\ref{fig:astrohist}, I over-plot a blue histogram with striped bars for the accepted source-lens pairs.
The filter on $\Delta\varpi/\sigma[\Delta\varpi]$
cleanly removes the peaks from the distributions of $\Delta\varpi/\sigma[\Delta\varpi]$,
$\Delta\mu_{\alpha*}/\sigma[\Delta\mu_{\alpha*}]$ and $\Delta\mu_{\delta}/\sigma[\Delta\mu_{\delta}]$.
A quick inspection of the S/N ratios of the total relative proper motions
for the accepted source-lens pairs reveals two pairs with S/N~<~3, which I also reject.
I now have 1,533 source-lens pairs remaining, with 1,257 unique lenses.

\subsubsection{Final Source-Lens Pair Selection}
\label{sec:finalsel}

Returning to the calculation of the improved maximum detection radius described in Section~\ref{sec:refsourcelenspairs},
I recompute the values of $\theta_{\mbox{\scriptsize det}}$ for the latest set of source-lens pairs using the
improved upper limit on the lens mass $M_{\mbox{\scriptsize max}}$ and adopting
$\Delta\varpi_{\mbox{\scriptsize max}}=\Delta\varpi+3\sigma[\Delta\varpi]$, which takes into account the
source parallax (when available). This modification is now possible because of the constraint applied to
$\Delta\varpi/\sigma[\Delta\varpi]$ in Section~\ref{sec:binary}. By rejecting all source-lens pairs
that do not approach each other to within an angular distance of less than $\theta_{\mbox{\scriptsize det}}$
during the maximal time baseline of the \textit{Gaia} mission,
I am left with a final sample that consists of 1,470 source-lens pairs, with 1,194 unique lenses.

\section{Threading The Needles}
\label{sec:events}

In Section~\ref{sec:finalsel}, I made a final selection of source-lens pairs from GDR2 that could potentially lead to microlensing
events by using conservative upper limits on the lens mass and the source-lens relative parallax, while also
considering their paths on the sky. However, to be able to predict a set of microlensing events and their properties from
these source-lens pairs, it is essential to have a reasonable estimate of the lens mass in each case.

\subsection{Lens Mass Estimates}
\label{sec:lens_masses}

\begin{figure*}
\centering
\epsfig{file=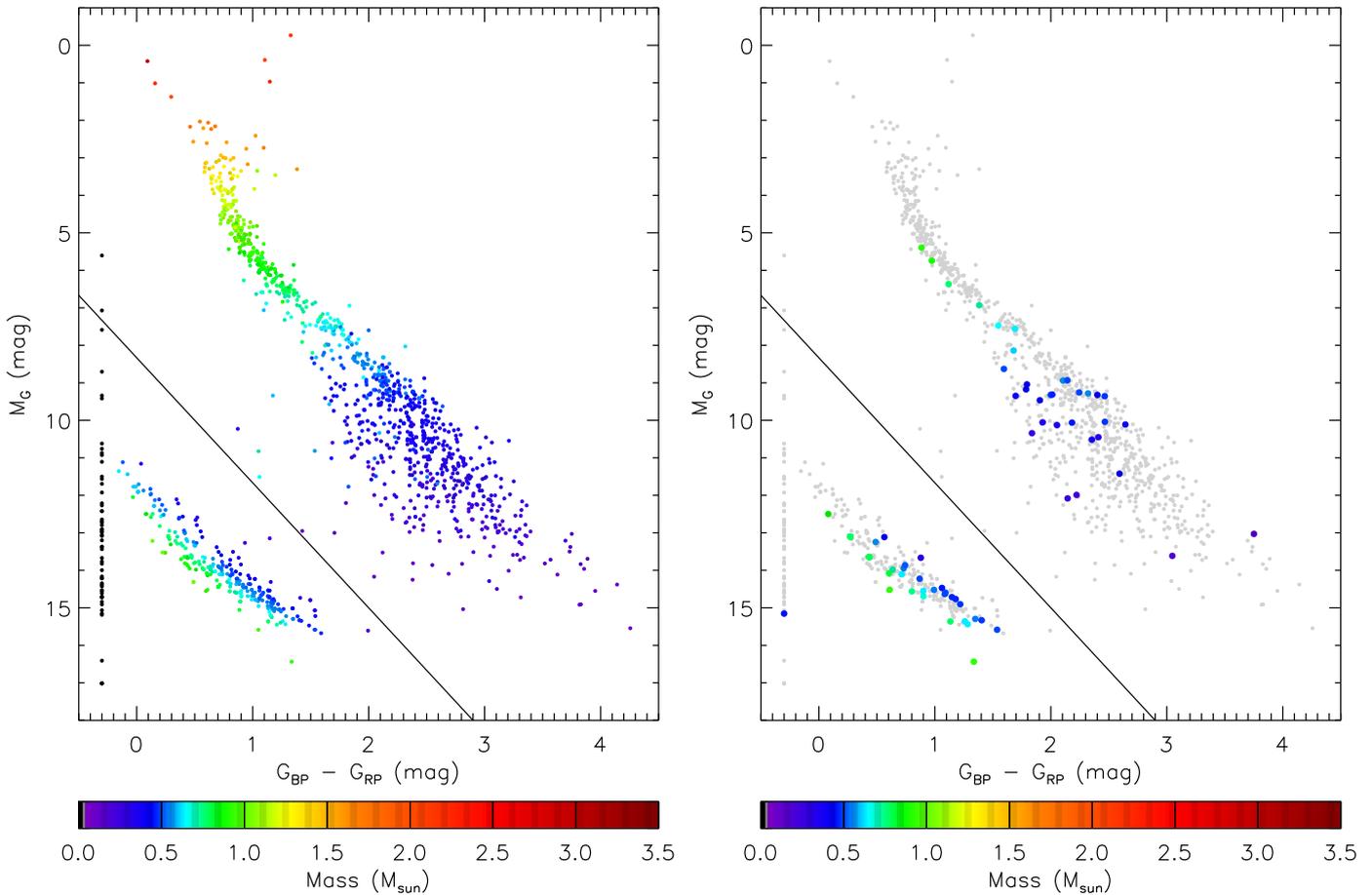,angle=0.0,width=\linewidth}
\caption{\textit{Left:} Hertzsprung-Russell diagram of $M_{G}$ versus $G_{\mbox{\scriptsize BP}}-G_{\mbox{\scriptsize RP}}$
         for the 1,194 lens stars in the final sample of source-lens pairs from Section~\ref{sec:finalsel}. The continuous
         black line joining (-1,5) and (5,25) mag separates white dwarf stars from main sequence dwarfs/subdwarfs
         (\citealt{kil2018}). The lens masses are indicated by the colour of the plot points (see the scale at the bottom
         of the panel). The 55 lenses that are present in 58 of the source-lens pairs that do not have
         $G_{\mbox{\scriptsize BP}}$ or $G_{\mbox{\scriptsize RP}}$ magnitudes in GDR2 are plotted with black points
         at fixed colour $G_{\mbox{\scriptsize BP}}-G_{\mbox{\scriptsize RP}}=-0.3$~mag and using their computed
         absolute magnitudes $M_{G}$. \textit{Right:} Same as the left-hand panel for the 64 lens stars in the 76 microlensing
         events found in Section~\ref{sec:findingevents}, except that the only lens that does not
         have a $G_{\mbox{\scriptsize BP}}$ or $G_{\mbox{\scriptsize RP}}$ magnitude is plotted in colour to indicate its estimated mass
         using photometry from external catalogues. The points from the left-hand panel are plotted in
         the background in light grey.
         \label{fig:hrd}}
\end{figure*}

In the left-hand panel of Figure~\ref{fig:hrd}, I plot absolute $G$-band magnitude $M_{G}$ against
$G_{\mbox{\scriptsize BP}}-G_{\mbox{\scriptsize RP}}$ colour (using {\tt PHOT\_BP\_MEAN\_MAG} and {\tt PHOT\_RP\_MEAN\_MAG})
for the 1,194 lens stars in the final sample of source-lens pairs from Section~\ref{sec:finalsel}.
The absolute $G$ magnitude is calculated from:
\begin{equation}
M_{G} = G + 5\log(\varpi_{\mbox{\scriptsize L}}) + 5
\label{eqn:absgmag}
\end{equation}
where $G$ is the apparent $G$-band mean magnitude ({\tt PHOT\_G\_MEAN\_MAG}). No attempt has been made to
account for reddening and extinction in this plot since $\sim$80\% of the lenses are at distances of less than 200~pc,
and the most distant lens is at $\sim$1.01~kpc. The plot is essentially a Hertzsprung-Russell diagram for the lens stars
and it reveals that the lens sample consists of main sequence and
white dwarf stars with a few stars on the giant branch. The main sequence lens stars are dominated by K and M dwarfs/subdwarfs
although their full range covers from late M dwarfs up to early A stars.
The white dwarf lens stars lie below the
continuous black line joining ($-$1,5) and (5,25) mag defined by \citet{kil2018}. Unfortunately, 55 lenses
that are present in 58 of the source-lens pairs do not have
$G_{\mbox{\scriptsize BP}}$ or $G_{\mbox{\scriptsize RP}}$ magnitudes in GDR2 and I plot them
in Figure~\ref{fig:hrd} at fixed colour $G_{\mbox{\scriptsize BP}}-G_{\mbox{\scriptsize RP}}=-0.3$~mag
using their computed absolute magnitudes $M_{G}$ (black points).

For the 1,075 main sequence and giant stars that lie above the continuous line in Figure~\ref{fig:hrd}, I use the
{\tt isochrones} Python package (\citealt{mor2015}) to estimate the lens masses. The package employs the MESA
(Paxton et al. 2011, 2013, 2015) Isochrones and Stellar Track Library
(MIST; \citealt{dot2016}; \citealt{cho2016}) and maximises the posterior probability of the fundamental
parameters (mass, age, metallicity, distance, and extinction) given the data. For each lens star,
I compute Sloan $g$ and $i$ magnitudes from $G$, $G_{\mbox{\scriptsize BP}}$ and $G_{\mbox{\scriptsize RP}}$
via the relevant transformations given in \citet{eva2018}, and I provide $\varpi_{\mbox{\scriptsize L}}$, $G$, $g$ and $i$,
along with their uncertainties, as input to the {\tt isochrones} package. I also
use the {\tt dustmaps} Python package with the Bayestar17 three-dimensional dust maps (Green et al. 2015, 2018)
to bound-above the extinction prior in {\tt isochrones} for each lens star where possible. The posterior
distributions are sampled using the MCMC ensemble sampler {\tt emcee} (\citealt{for2013}) with 300 walkers.
Each walker executes a burn-in of 300 steps, and then iterates through 500 subsequent steps of which the last 100 steps are
recorded. I adopt the median of the posterior sample as the estimate of the lens mass in each case.
The scatter in the lens mass posterior samples is typically $\sim$2-10\%, which is comparable to the
systematic errors in the MESA evolutionary tracks ($\sim$5-10\% differences with other evolutionary track libraries; \citealt{pax2011}).
In fact, MIST isochrones are known to predict colours that are too blue for stars with masses below
$\sim$0.6-0.7$M_{\sun}$ (\citealt{cho2016}), which implies that the mass estimates for the lower mass stars may be 
systematically underestimated. Hence the estimated masses may be in error by up to $\sim$15\% in the worst cases
with a tendency towards underestimation. However, these errors are too small to substantially affect the microlensing predictions
presented in this paper.
The lens masses estimated by this method are indicated in Figure~\ref{fig:hrd} by the colours of the plot points.

There are 337 white dwarf lenses below the continuous line in Figure~\ref{fig:hrd} with distances ranging
from $\sim$4.3 to 237~pc. To estimate their masses, I interpolate the evolutionary cooling sequences
for DA- and DB-type white dwarfs computed specifically for the \textit{Gaia} passbands (Pierre Bergeron - private communication;
\citealt{hol2006}; \citealt{kow2006}; \citealt{tre2011}; \citealt{ber2011}). The mass estimates from the DA and DB
cooling sequences are always very similar (to within $\sim$1-15\%) and for the purposes of predicting microlensing events,
these differences are unimportant. Therefore, for each white dwarf lens star, I adopt the greater of the two mass estimates.
Again, in Figure~\ref{fig:hrd}, the lens masses estimated by this method are indicated by the colours of the plot points.

For the 55 lens stars that do not have $G_{\mbox{\scriptsize BP}}$ or $G_{\mbox{\scriptsize RP}}$ magnitudes in GDR2,
I continue to adopt for the moment the improved upper limit on the lens mass $M_{\mbox{\scriptsize max}}$
computed in Section~\ref{sec:refsourcelenspairs}.

\subsection{Finding Microlensing Events}
\label{sec:findingevents}

I now have all of the ingredients necessary for each source-lens pair to be able to predict microlensing
events and their properties. For each of the 1,470 source-lens pairs from Section~\ref{sec:finalsel}, I perform
1,000 Monte Carlo simulations of the source and lens paths on the sky. Each simulation is generated
using the following procedure:
\begin{enumerate}[(i)]
\item I draw a set of astrometric parameters for the lens star
      from a multi-variate Gaussian distribution defined by the lens astrometric solution parameter values and their covariance
      matrix provided in GDR2. I do the same for the source star.
\item I calculate the Einstein radius $\theta_{\mbox{\scriptsize E}}$ (Equation~\ref{eqn:einsteinradius})
      using the lens mass estimate from Section~\ref{sec:lens_masses},
      and the lens and source parallaxes $\varpi_{\mbox{\scriptsize L}}$ and $\varpi_{\mbox{\scriptsize S}}$, respectively,
      drawn in step~(i).
\item I compute the path of the source relative to the lens (Section~\ref{sec:starpaths}) in units of normalised source-lens separation $u$ for the
      time period from $t=2456863.5$~d to $t=2461246.5$~d (Section~\ref{sec:times}).
\item I adopt \textit{Gaia}'s resolution of 103~mas and I calculate the maximum change in each of the six microlensing
      signals $A$, $\delta_{\mbox{\scriptsize mic}}$, $A_{1}$, $\theta_{2}$, $A_{\mbox{\scriptsize LI}_{2}}$, and
      $\theta_{\mbox{\scriptsize LI}_{2}}$ over the 12-year time baseline. To do this, I use the lens-to-source flux ratio
      $f_{\mbox{\scriptsize L}} / f_{\mbox{\scriptsize S}}$ derived from the \textit{Gaia} $G$-band photometry
      and the relevant equations from Section~\ref{sec:microastro}.
      I refer to these values as ``delta'' microlensing signals.
\end{enumerate}
I then calculate the median of each of the delta microlensing signals over all of the simulations for
the source-lens pair. Collecting these results for the 1,470 source-lens pairs, I reject
all source-lens pairs for which none of the median delta microlensing signals exceed 0.4~mmag for photometric signals or
0.131~mas for astrometric signals (Section~\ref{sec:refsourcelenspairs}).

For the 55 lens stars in 58 source-lens pairs that do not have $G_{\mbox{\scriptsize BP}}$ or $G_{\mbox{\scriptsize RP}}$ magnitudes in GDR2,
only 16 lens stars in 16 source-lens pairs remain after this procedure. For these 16 lenses, I looked up their counterparts in the PPMXL
(\citealt{roe2010}) and Pan-STARRS1 (\citealt{cha2016}) catalogues, and I used the photometry from these catalogues,
combined with their GDR2 parallaxes and $G$ magnitudes, to estimate their masses using the methods described in
Section~\ref{sec:lens_masses} (the 16 lenses consist of 15 main sequence stars and one white dwarf). I then re-ran the above procedure,
which left only one of these lenses in a single source-lens pair that exhibits a detectable delta microlensing signal
(microlensing event ME28; Section~\ref{sec:astromevents}).

The final set of predicted microlensing events consists of 76 events caused by 64 unique lens stars. I name these events ME1-ME76. One lens causes nine events
(ME1-ME9; Section~\ref{sec:lawd37}), while another lens causes five events (ME10-ME14; Section~\ref{sec:stein2051b}). The remaining 62 events are caused by 62 unique lens stars.
The 64 lenses are at distances ranging from $\sim$3.29 to 366~pc. The lenses are typically considerably brighter than the sources
with differences ranging from $\sim$12.2 mag brighter to $\sim$1.8 mag fainter. The lens proper motions range from
$\sim$23.1~mas/year to 6.896~arcsec/year. In the right-hand panel of Figure~\ref{fig:hrd}, I plot the Hertzsprung-Russell
diagram for the lenses in these microlensing events, with the lenses from the left-hand panel plotted in the background in
light grey. The lenses break down into 34 main sequence stars and 30 white dwarf stars.

For each microlensing event, I use the results of the Monte Carlo simulations to calculate the median values of the
event properties $\theta_{\mbox{\scriptsize E}}$, $u_{0}$, and $t_{0}$, and of all of the delta microlensing signals.
I also calculate the median full-width duration at half-maximum signal for each signal.
I compute 1-sigma confidence intervals for each quantity, which may be asymmetric. The results
are reported in Tables~\ref{tab:lawd37}~to~\ref{tab:remainingevents}. Asymmetric uncertainties are only reported where necessary.

\section{Sewing The Future} 
\label{sec:results}

In this section, I present and discuss the 76 microlensing events that have been identified. Nine events are gold
events in the sense that they will exhibit both photometric and astrometric signals above the precision limits discussed in Section~\ref{sec:refsourcelenspairs}.
The 67 remaining events will exhibit purely astrometric signals, with 12 of these giving rise to astrometric shifts of more than
0.5~mas.

\subsection{Nine Microlensing Events Caused By LAWD~37}
\label{sec:lawd37}

\begin{figure*}
\centering
\epsfig{file=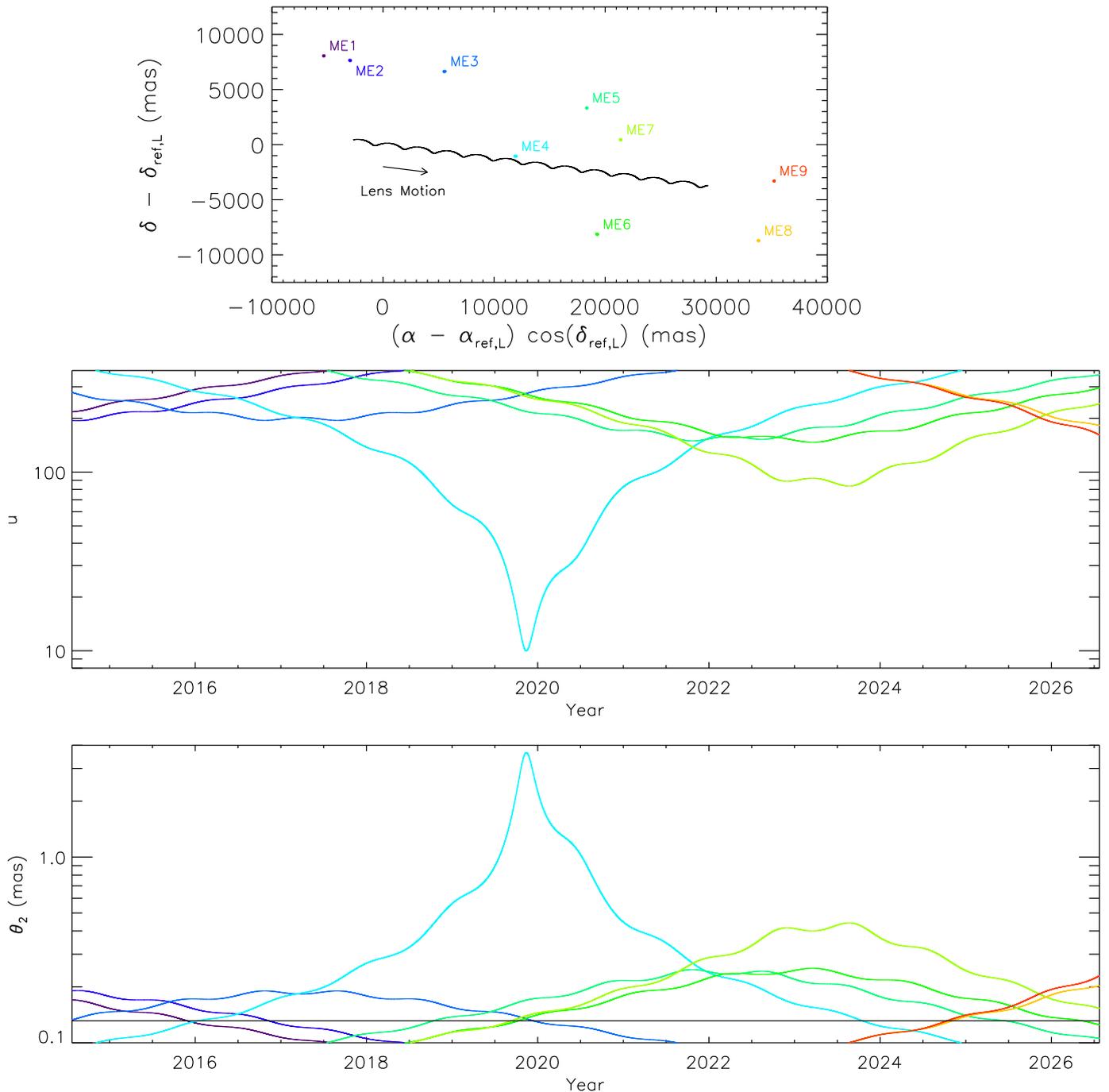,angle=0.0,width=\linewidth}
\caption{\textit{Top:} Path on the sky of the white dwarf lens star LAWD~37 (WD~1142$-$645), and nine source stars in the microlensing
                       events ME1-ME9, over the time baseline of an extended \textit{Gaia} mission. The coordinate axes are measured
                       relative to $(\alpha_{\mbox{\scriptsize ref,L}},\delta_{\mbox{\scriptsize ref,L}})$.
         \textit{Middle:} Time-evolution of the normalised source-lens separation $u$ for each microlensing event.
         \textit{Bottom:} Time-evolution of the deflection $\theta_{2}$ (mas) of the major source image.
                          The horizontal black line indicates the astrometric precision limit of 0.131~mas from Section~\ref{sec:refsourcelenspairs}.
         \label{fig:lawd37}}
\end{figure*}

The microlensing event presented by M18, and discussed in Sections~\ref{sec:intro}~and~\ref{sec:examplegaia}, is the event
ME4. The improved parameters of the astrometric solutions from GDR2 for the white dwarf lens star LAWD~37 and the source star yield a refined prediction
of their separation at closest approach of $\sim$368.1$\pm$1.6~mas which will occur on 14th November 2019 at 5~h ($\pm$3~h).
The mass that I estimate for LAWD~37 in Section~\ref{sec:lens_masses} is $\sim$28\% larger than that adopted by M18, which means that
$\theta_{\mbox{\scriptsize E}}\approx36.97\pm0.02$~mas and $u_{0}\approx9.95\pm0.05$ are larger and smaller, respectively, than their
corresponding values predicted by M18. For the same reason, I also predict a larger change in the
deflection of the source position of $\Delta\theta_{2}\approx3.600\pm0.018$~mas over the 12-year baseline. The
event has a full-width duration at half-maximum signal of $\sim$122.2$\pm$0.6~d.

These results support the conclusion that the astrometric microlensing signal can be detected by \textit{Gaia} for the most favourable scanning angles
(M18; Section~\ref{sec:examplegaia} this paper). However, I note that the astrometric noise model for \textit{Gaia} observations in M18 is considerably
more optimistic than that of R18 adopted here (Section~\ref{sec:astroprec}). M18 quote $\sigma_{\mbox{\scriptsize AL}} \approx 0.2/\sqrt{9} \approx 0.067$~mas 
for the $G\approx18.6$~mag source star whereas Equation~\ref{eqn:modelastromal} yields $\sigma_{\mbox{\scriptsize AL}} \approx 1.17$~mas.
Clearly, if the R18 astrometric noise model is correct, then the astrometric signal will only be revealed at $\ga1\sigma_{\mbox{\scriptsize AL}}$
for a small range of scan angles, and it will certainly be impossible for \textit{Gaia} to measure the mass of LAWD~37 to within $\sim$3\% as claimed by M18.
The \textit{HST} has a much better outlook for observing this event because the observational setup can be optimised for the lens and source properties.

ME4 is not the only microlensing event to be caused by LAWD~37 between 2014 and 2026, although it does have the largest peak amplitude. In total there are nine astrometric microlensing 
events ME1-ME9 that will all unfold completely in the partially-resolved regime (Table~\ref{tab:lawd37} and Figure~\ref{fig:lawd37}). Events ME1-ME3 
are low-amplitude events with $\Delta\theta_{2}\approx0.13$-0.15~mas that are already, or currently, finishing. Apart from ME4, the most promising
remaining event is ME7\footnote{The source stars for ME7 and ME8 have similar proper motions and parallaxes which suggests that they are
a pair of binary stars.} which will achieve $\Delta\theta_{2}\approx0.386\pm0.004$~mas, although it will unfold slowly between 2020 and 2026. All of these
events overlap in time and they highlight the fact that when performing astrometry on the stars in the near-field of LAWD~37, the lensing effect
will need to be modelled as part of the astrometric solution. However, this approach has the advantage that multiple stars can be used simultaneously to
measure the mass of LAWD~37.

\subsection{Five Microlensing Events Caused By Stein~2051~B}
\label{sec:stein2051b}

\begin{figure*}
\centering
\epsfig{file=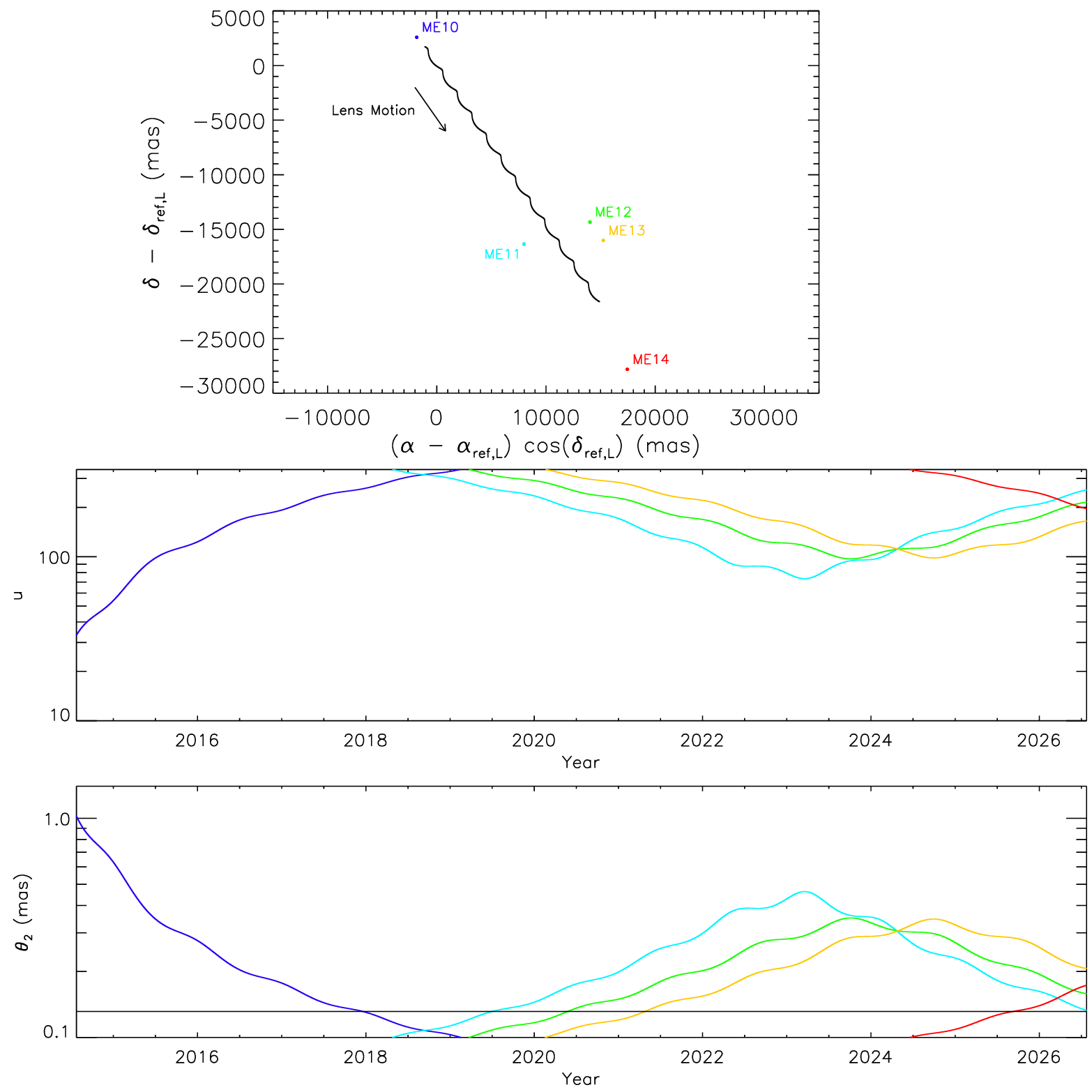,angle=0.0,width=\linewidth}
\caption{\textit{Top:} Path on the sky of the white dwarf lens star Stein~2051~B (WD~0426+588), and five source stars in the microlensing
                       events ME10-ME14, over the time baseline of an extended \textit{Gaia} mission. The coordinate axes are measured
                       relative to $(\alpha_{\mbox{\scriptsize ref,L}},\delta_{\mbox{\scriptsize ref,L}})$.
         \textit{Middle:} Time-evolution of the normalised source-lens separation $u$ for each microlensing event.
         \textit{Bottom:} Time-evolution of the deflection $\theta_{2}$ (mas) of the major source image. 
                          The horizontal black line indicates the astrometric precision limit of 0.131~mas from Section~\ref{sec:refsourcelenspairs}.
         \label{fig:stein2051b}}
\end{figure*}

ME10 happens to be the tail-end of the microlensing event observed by \citet{sah2017} for the white dwarf lens star Stein~2051~B (WD~0426+588). The lens is part of a binary
system where the other component Stein~2051~A is at a separation of $\sim$10.22\arcsec. However, the binary
orbital motion ($P\ga1000$~years) is too slow to affect the microlensing predictions presented here (\citealt{hei1990}).
Stein~2051~B will cause four more astrometric microlensing events ME11-M14 between 2020 and 2026 that will all unfold completely in the partially-resolved
regime (Table~\ref{tab:stein2051b} and Figure~\ref{fig:stein2051b}). Three of these events will reach peak signals of $\Delta\theta_{2}\approx0.3$~mas and above in relatively quick succession.
Unfortunately, \textit{Gaia} will not be able to detect any of them, even with the observations already acquired of the tail-end of ME10, because $\sigma_{\mbox{\scriptsize AL}}$
is too large for such faint source stars. However, \textit{HST} will be able to repeat what it has already achieved for ME10, and observations
of multiple lensed source stars will help to further constrain the mass of Stein~2051~B.

\subsection{Nine Photometric Microlensing Events}
\label{sec:photevents}

\begin{figure*}
\centering
\epsfig{file=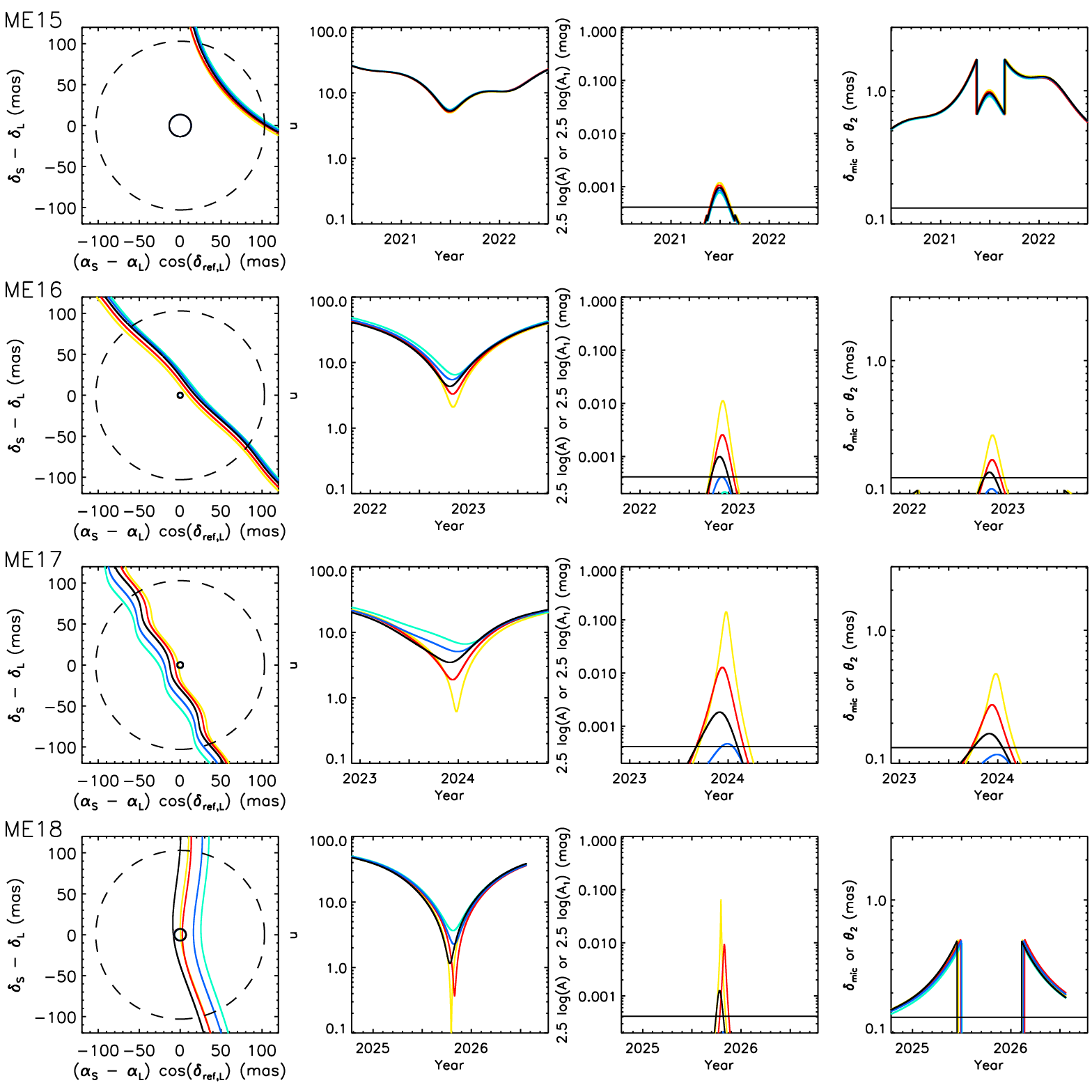,angle=0.0,width=\linewidth}
\caption{Photometric microlensing events ME15-ME18.
         In all panels, five curves are plotted with the colours yellow, red, black, blue, and cyan. Each curve corresponds to the 2.3, 15.9, 50, 84.1,
         and 97.7 percentiles, respectively, of the results of the Monte Carlo simulations performed in Section~\ref{sec:findingevents} after they have been ordered
         by increasing $u_{0}$. The yellow and cyan curves are plotted first, followed by the red and blue curves, and finally the black curve, which is why
         the black curve is the most visible when the individual curves are hard to distinguish.
         \textit{Left-hand panels:} Path of the source star relative to the lens star. The Einstein ring is shown as a circle of radius $\theta_{\mbox{\scriptsize E}}$
                                    centred on the lens position (also plotted five times with five different colours). The resolution of \textit{Gaia} is indicated as a circle
                                    of radius 103~mas centred on the lens position (dashed curve).
         \textit{Middle left-hand panels:} Time-evolution of the normalised source-lens separation $u$.
         \textit{Middle right-hand panels:} Time-evolution of the photometric signals $2.5\log(A)$ (mag; unresolved regime) and $2.5\log(A_{1})$ (mag; partially-resolved regime).
                                            The horizontal black line indicates the photometric precision limit of 0.4~mmag from Section~\ref{sec:refsourcelenspairs}.
         \textit{Right-hand panels:} Time-evolution of the astrometric signals $\delta_{\mbox{\scriptsize mic}}$ (mas; unresolved regime) and $\theta_{2}$ (mas; partially-resolved regime).
                                     The horizontal black line indicates the astrometric precision limit of 0.131~mas from Section~\ref{sec:refsourcelenspairs}.
         \label{fig:photevents1}}
\end{figure*}

\begin{figure*}
\centering
\epsfig{file=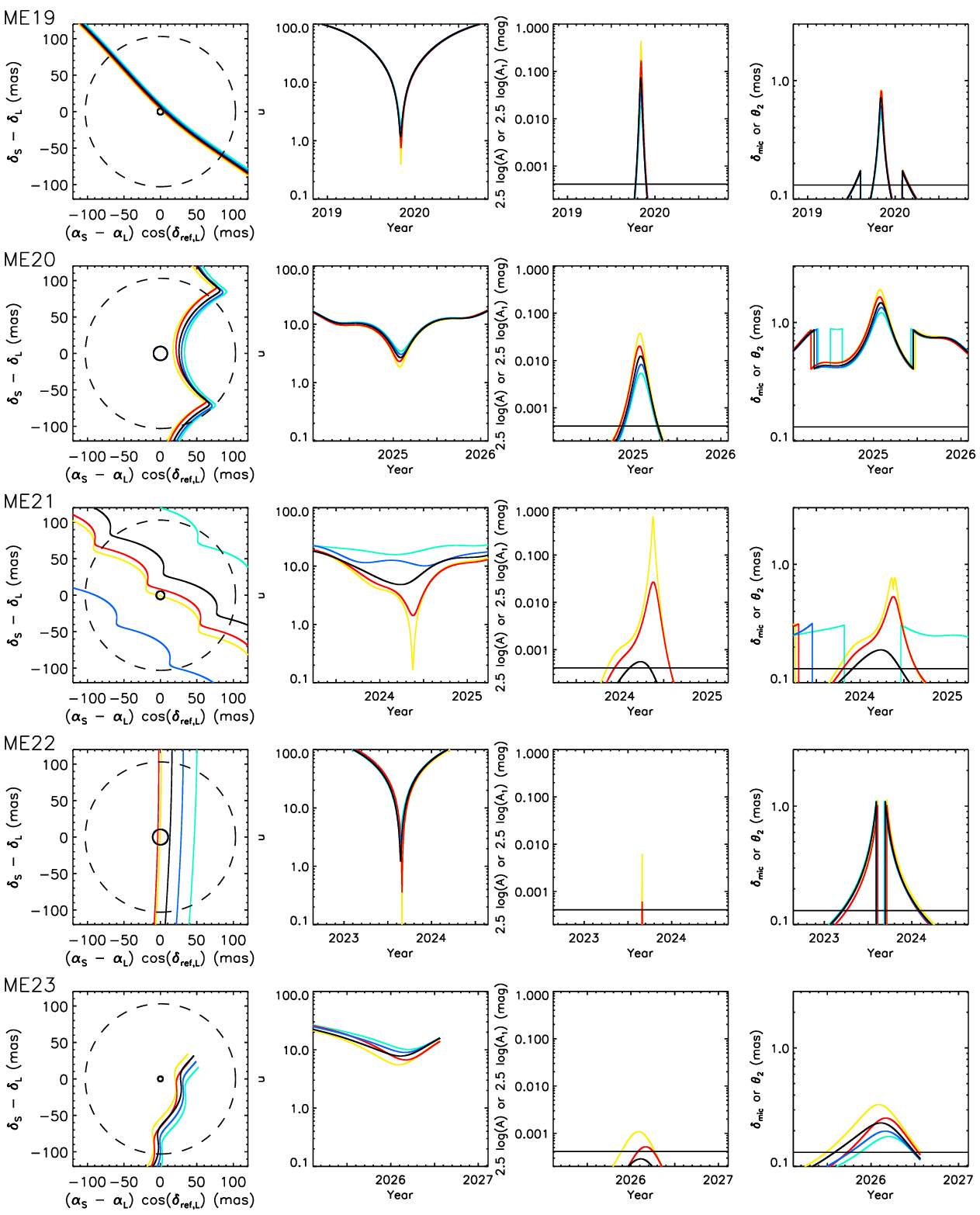,angle=0.0,width=\linewidth}
\caption{Photometric microlensing events ME19-ME23. The format of the figure is the same as in Figure~\ref{fig:photevents1}.
         \label{fig:photevents2}}
\end{figure*}

The nine events ME15-ME23 are predicted to yield both photometric and astrometric signals as they unfold (Tables~\ref{tab:photevents1}~and~\ref{tab:photevents2}),
and all of them will unfold in both the partially-resolved and unresolved regimes when considering the resolution of \textit{Gaia} (Figures~\ref{fig:photevents1}~and~\ref{fig:photevents2}).
ME18, ME19 and ME22 each have better than a $\sim$33\% probability that the source will pass within the Einstein ring of the lens.
However, the prediction of the exact amplitude of the photometric signal that will be observed is quite uncertain in most cases simply because the uncertainties on the relative source-lens
motions are similar to, or larger than, the Einstein radius. Specifically, ME17-ME19, ME21 and ME22 could potentially evolve into high-magnification microlensing events.
In the unresolved regime, both the photometric and astrometric signals can be boosted by observing the event in a passband for which
$f_{\mbox{\scriptsize L}} / f_{\mbox{\scriptsize S}}$ is minimised (Equations~\ref{eqn:magnification}~and~\ref{eqn:centroidshift}).
The \textit{Gaia} $G$-band used in this analysis is sub-optimal for most events and a tailored follow-up observing strategy can greatly improve the observed
signals.

Assessing whether \textit{Gaia} will be able to detect the photometric signals of any of the microlensing events presented in this section is very difficult because
of the satellite's complicated scanning law. The durations of the photometric signals range from $\sim$7-133~d, and \textit{Gaia} returns to the same field every $\sim$63~d on average,
which implies that the photometric events can easily be missed. It is much more advisable to target these events with an appropriate telescope during the predicted event
duration in order to characterise them properly. With regards to the accompanying astrometric signals, \textit{Gaia} can only detect ME15, ME19 and ME20 at $\sim$2-4$\sigma_{\mbox{\scriptsize AL}}$ (i.e.
for the most favourable scanning angles).

\textit{ME15:} The lens star OGLE~SMC115.5.319 is a high proper motion white dwarf star (\citealt{pol2011}). It lies in the direction of the Small Magellanic Cloud (SMC)
and it is a member of a binary system, the other component of which is the M5 dwarf OGLE~SMC115.5.12 (separation $\sim$13.10\arcsec; $P\ga7300$~years).
The lens has $G\approx17.65$~mag which is only $\sim$0.5~mag brighter than the source star.
The event will peak on 30th June 2021 at 5~h ($\pm$1~d) with a photometric amplitude of $\sim$1~mmag and it will achieve a maximum source deflection of $\sim$1.64~mas (detectable
by \textit{Gaia} at $\sim2\sigma_{\mbox{\scriptsize AL}}$). In Figure~\ref{fig:photevents1},
the jumps in the astrometric curve are due to the event switching between partially-resolved and unresolved microlensing at the \textit{Gaia} resolution
of 103~mas. The lens flux suppresses both the photometric and astrometric signals in the unresolved regime. Observing in $G_{\mbox{\scriptsize BP}}$ would boost
the peak photometric signal by a factor of $\sim$1.3.

\textit{ME16:} The lens star SDSS~J035037.54+112707.9 is a spectroscopically confirmed M2 dwarf star (\citealt{wes2011}) at a distance of $\sim$366~pc (the most distant lens star in my sample).
The photometric signal is somewhat diluted by the lens star, but it may still peak at $\sim$0.01~mag (+2$\sigma$) on 31st October 2022 ($\pm$7~d).

\textit{ME17:} Peaking on 27th December 2023 ($\pm$20~d), this event will be caused by an M dwarf lens star (not in SIMBAD\footnote{\url{http://simbad.u-strasbg.fr/simbad/}}) and it may
reach a peak magnification of $\sim$0.15~mag (+2$\sigma$). This is the second most distant lens star in my sample ($\sim$270~pc).

\textit{ME18:} The lens star G192-52 is a spectroscopically confirmed M1 subdwarf (\citealt{bai2016}) that is part of a visual binary (\citealt{mas2001}; binary separation of $\sim$596\arcsec),
and it is $\sim$6~mag brighter than the source star. Even though the photometric signal will be highly suppressed by the lens blend flux, there is still a resonable chance that the
photometric signal will peak at above $\sim$0.06~mag (+2$\sigma$) on 23rd October 2025 ($\pm$15~d) since the source has a $\sim$42\% probability of passing within the Einstein ring.
Observing in $G_{\mbox{\scriptsize BP}}$ would boost the peak photometric signal by a factor of $\sim$3.6.

\textit{ME19:} This is by far the most promising photometric microlensing event presented here. The source has a $\sim$33\% probability of passing within the Einstein ring of
the M dwarf lens star and the $\pm$1-sigma range for the photometric magnification is between $\sim$0.039-0.158~mag, with the possibility of reaching above $\sim$0.45~mag (+2$\sigma$).
The event is favourable because the lens star at $G\approx17.18$~mag is only $\sim$1.0~mag brighter than the source star, it will have a relatively short duration of
$\sim$7.7~d, and the peak is well-constrained to occur on 3rd November 2019 at 15~h ($\pm$32~h).
Furthermore, the astrometric signal is detectable by \textit{Gaia} at $\sim2\sigma_{\mbox{\scriptsize AL}}$.
Observing in $G_{\mbox{\scriptsize BP}}$ would boost the peak signals by a factor of $\sim$1.3.

\textit{ME20:} The lens star is a white dwarf (not in SIMBAD) at $\sim$40.6~pc that is approximately the same brightness as the source star. The photometric magnification at peak is well-constrained
to lie in the $\pm$2-sigma range $\sim$0.006-0.04~mag which is easily detectable from both ground- and space-based telescopes. The peak will occur on 31st January 2025 ($\pm$3~d) and the photometric
event will last $\sim$43~d. The astrometric event will be much slower and it is detectable by \textit{Gaia} at $\sim4\sigma_{\mbox{\scriptsize AL}}$.

\textit{ME21:} The source has positional uncertainties of $\sim$29~mas in each coordinate direction, which is $\sim$5 times larger than the Einstein radius. Consequently the source
path relative to that of the lens is rather uncertain. Unfortunately, this makes any predictions for this event unreliable (e.g. the photometric signal can be anything from below 0.4~mmag to 1~mag).
The lens is a low-mass white dwarf star (not in SIMBAD), and the event will peak during April 2024 ($\pm$2~months).

\textit{ME22:} The lens star Wolf~851 is a spectroscopically confirmed K7 subdwarf (\citealt{giz1997}). The source has a $\sim$40\%  probability of passing within the Einstein ring.
However, since the lens is $\sim$9~mag brighter than the source, the photometric signal will be highly suppressed and it may only reach $\sim$6~mmag (+2$\sigma$). The peak will
occur on 28th August 2023 ($\pm$4~d).
Observing in $G_{\mbox{\scriptsize RP}}$ would boost the peak photometric signal by a factor of $\sim$1.5.

\textit{ME23:} The M dwarf lens star (not in SIMBAD) at $G\approx18.2$~mag is of approximately the same brightness as the source. Since the event will happen near the end of the time period considered,
the source path relative to that of the lens is somewhat uncertain, and the peak photometric signal of $\sim$1~mmag (+2$\sigma$) is predicted to occur on 28th February 2026 ($\pm$12~d).

\subsection{Ten Astrometric Microlensing Events}
\label{sec:astromevents}

\begin{figure*}
\centering
\epsfig{file=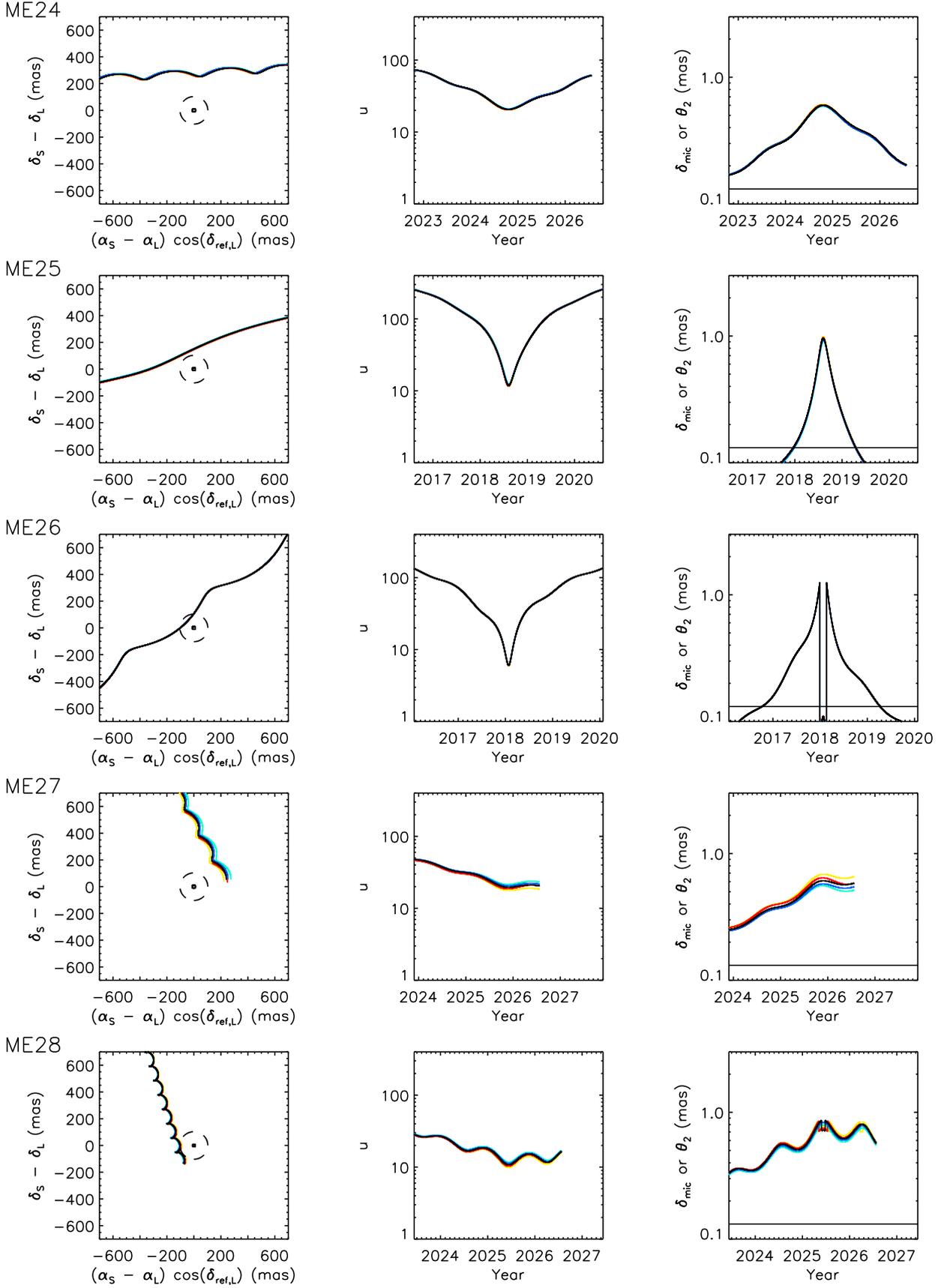,angle=0.0,width=\linewidth}
\caption{Astrometric microlensing events ME24-ME28. The format of the figure is the same as in Figure~\ref{fig:photevents1} except that the
         panels for the non-existent photometric signals have been dropped.
         \label{fig:astromevents1}}
\end{figure*}

\begin{figure*}
\centering
\epsfig{file=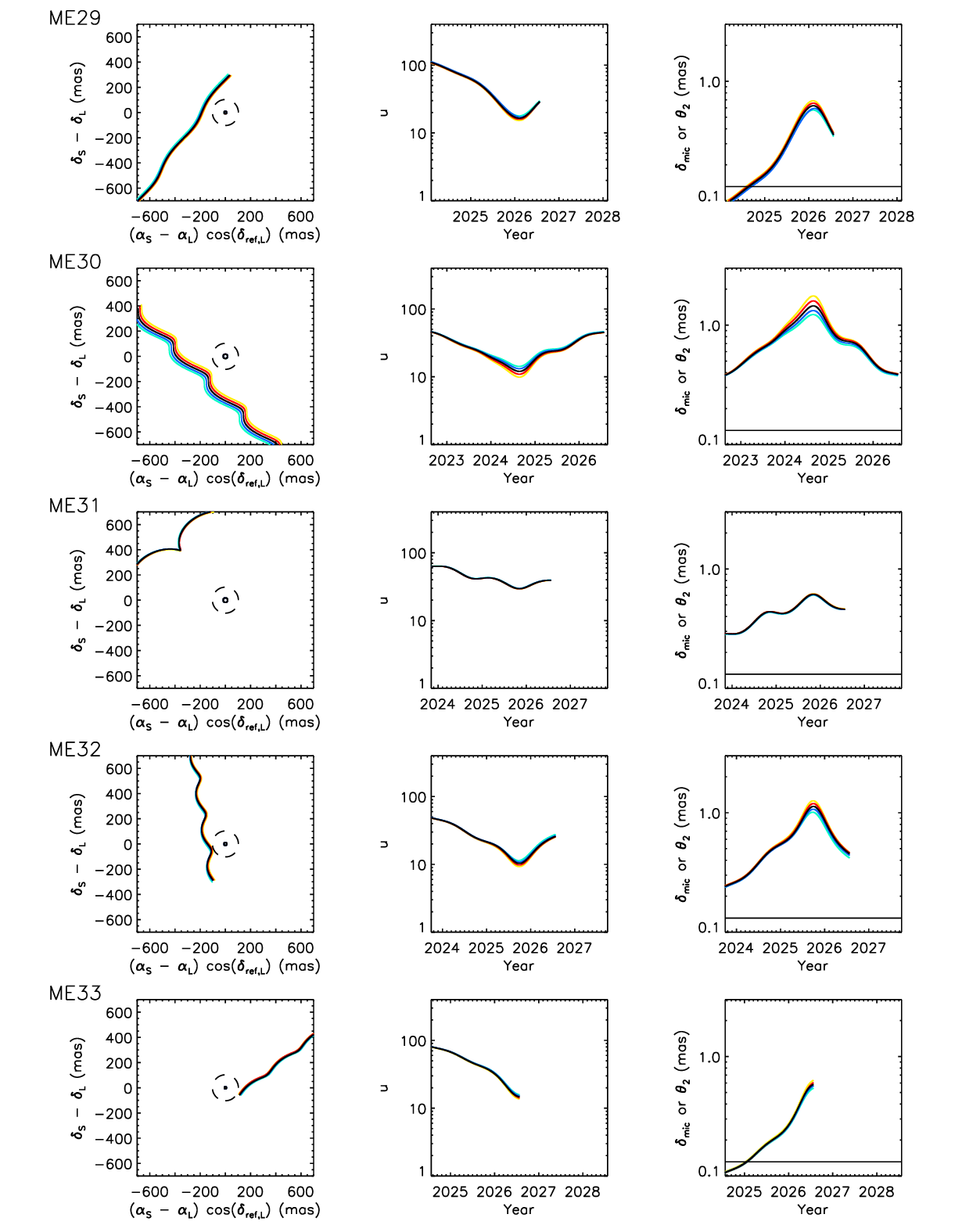,angle=0.0,width=\linewidth}
\caption{Astrometric microlensing events ME29-ME33. The format of the figure is the same as in Figure~\ref{fig:photevents1} except that the
         panels for the non-existent photometric signals have been dropped.
         \label{fig:astromevents2}}
\end{figure*}

The ten events ME24-ME33 will exhibit astrometric signals as they unfold (Tables~\ref{tab:astromevents1}~and~\ref{tab:astromevents2}). However, their predicted photometric signals
are too small to be significant because $u_{0}>6$ for all of these events. ME26 and ME28 will unfold in both the partially-resolved and unresolved regimes when considering
the resolution of \textit{Gaia} (Figures~\ref{fig:astromevents1}~and~\ref{fig:astromevents2}), while the other events will all unfold completely in the partially-resolved
regime. \textit{Gaia} will be able to detect ME26, ME28, ME31 and ME33 at $\sim$11, 3, 1 and 1$\sigma_{\mbox{\scriptsize AL}}$, respectively (i.e. for the most favourable scanning angles),
while the remaining events will be undetectable. However, all of the events presented in this section are within the detection capabilities of \textit{HST}.

\textit{ME24:} The white dwarf lens star LSPM~J0024+6834N (\citealt{lep2005}) at $\sim$36.1~pc has a brighter visual companion (G~242-54; \textit{Gaia} source ID 529594417069407744; M dwarf)
with similar proper motion and parallax at a separation of 2.34\arcsec,
which corresponds to a projected orbital separation of $\sim$84~au. Fortunately, the orbital motion ($P\ga800$~years) is too slow to affect the microlensing prediction presented here. The event
will peak on 19th October 2024 at 10~h ($\pm$25~h).

\textit{ME25:} The M3 dwarf lens star Ross~322 (\citealt{bid1985}) will cause a relatively fast-unfolding (duration $\sim$124~d) astrometric microlensing event during mid-2018, peaking
on 10th August 2018 at 7~h ($\pm$17~h). The signal will reach nearly 1~mas and the event is a great target for observations starting immediately.

\textit{ME26:} Unfortunately this event in the direction of the SMC is currently declining, although \textit{Gaia} will already have taken useful observations for which the astrometric signal is detectable at
$\sim$11$\sigma_{\mbox{\scriptsize AL}}$. Due to the brightness difference between the lens (M3-4 dwarf; \citealt{mas2003}) and the source, the signal is only significant in the partially-resolved microlensing
regime, with a maximum source deflection of $\sim$1.2~mas. The peak occurred on 25th January 2018 at 15~h ($\pm$2~h) and by mid-2018 the signal will have dropped to $\sim$0.3~mas.
Observations are encouraged immediately to catch this event before the signal becomes too weak.

\textit{ME27:} The lens star is a white dwarf (not in SIMBAD) that will cause an astrometric signal of $\sim$0.55~mas on 25th November 2025 ($\pm$5~d).

\textit{ME28:} This is one of only two events in the sample of 76 microlensing events for which the source is significantly brighter than the lens
($f_{\mbox{\scriptsize L}} / f_{\mbox{\scriptsize S}}\approx0.20$). Unfortunately,
in this case, the lens does not have a $G_{\mbox{\scriptsize BP}}$ or $G_{\mbox{\scriptsize RP}}$ magnitude in GDR2. However, using data from the PPMXL catalogue (\citealt{roe2010}), I was
able to surmise that this is a $\sim0.46M_{\sun}$ M-dwarf star (not in SIMBAD). Interestingly, the photometric signal $A_{\mbox{\scriptsize LI}_{2}}$ will nearly reach 0.4~mmag
in the partially-resolved regime. Observing this event in passbands where $f_{\mbox{\scriptsize L}} / f_{\mbox{\scriptsize S}}$ is even smaller than $\sim$0.20 will potentially
enable the first detection of the minor source image in a microlensing event in our galaxy (Equation~\ref{eqn:ALI2}). The event will peak on 12th June 2025 ($\pm$2~d).

\textit{ME29:} The peak of this event will occur towards the end of the \textit{Gaia} extended mission on 5th February 2026 ($\pm$4~d) with an astrometric signal
of $\sim$0.61~mas. The lens star Ross~528 is a spectroscopically confirmed K4 dwarf (\citealt{bid1985}).

\textit{ME30:} The lens star LAWD~66 (WD~1708$-$147) is a spectroscopically confirmed DQ6 white dwarf (\citealt{mcc1999}) at a distance of $\sim$22.66~pc.
The event will peak on 23rd August 2024 at 10~h ($\pm$15~h) with a peak signal of $\sim$1.4~mas. Unfortunately the source is rather faint at $G\approx20.8$~mag.

\textit{ME31:} The nearby ($\sim$13.51~pc) spectroscopically confirmed DC-type white dwarf lens WD~1743$-$545 (\citealt{sub2017})
will cause a peak astrometric signal of $\sim$0.55~mas on 5th November 2025 at 7~h ($\pm$1~h).
The time of closest approach is very well constrained because of the form of the path taken by the source relative to the lens.

\textit{ME32:} On 26th September 2025 ($\pm$4~d), the event caused by the spectroscopically confirmed DA-type white dwarf lens LSPM~J1913+2949 (\citealt{lim2013})
will peak with an astrometric signal of $\sim$1.1~mas.

\textit{ME33:} This event will still be rising towards its peak at the end of the time period considered in this paper (25th July 2026). The peak signal will be a little
above $\sim$0.56~mas. The lens star Ross~213 is a spectroscopically confirmed K5 dwarf (\citealt{bid1985}).

\subsection{Low-Amplitude Astrometric Microlensing Events}
\label{sec:astromevents_low}

The 43 astrometric microlensing events ME34-ME76 will give rise to peak astrometric signals with amplitudes in the range 0.131-0.5~mas.
I present these low-amplitude astrometric events in Table~\ref{tab:remainingevents} (provided electronically) for exploration by the
astronomical community. No further analysis is performed here.

\section{Summary And Conclusions}
\label{sec:conc}

I selected 1,470 source-lens pairs, with 1,194 unique lenses, from GDR2 that could potentially lead to microlensing events during the time baseline of an
extended \textit{Gaia} mission lasting from 25th July 2014 until 25th July 2026. Analysing these pairs in detail, I found that 76 microlensing events caused by
64 unique lenses will occur that exhibit microlensing signals exceeding the precision limits of 0.4~mmag (photometric) and 0.131~mas (astrometric).
Nine and five astrometric microlensing events will be caused by LAWD~37 and Stein~2051~B, respectively. These lens stars are very nearby white dwarfs passing through relatively crowded fields
which explains their efficiency at producing microlensing events.

A further nine events will yield both photometric and astrometric signals. These are highly desirable events for follow-up
observations at the correct moment since they have the potential to constrain the lens mass much better than events that only exhibit an astrometric signal.
Five of these photometric events, ME17-ME19, ME21 and ME22 could potentially achieve high-magnification  which can be used to probe lens binarity or for planetary companions.
Specifically, ME19, which will peak at the beginning of November 2019, is the most promising microlensing event presented in this paper, and its characterisation
will allow a precise measurement of the mass of a late-type M dwarf.

Ten microlensing events will yield astrometric signals with amplitudes above 0.5~mas, and two of these events (ME25 and ME26) are ongoing during 2018, requiring immediate observation.
The 43 remaining events are all low-amplitude astrometric events with amplitudes between 0.131 and 0.5~mas reported for further exploration by the astronomical community.

The \textit{Gaia} satellite has ushered in an era of being able to reliably predict gravitational microlensing events for (at least) the next decade.
The third data release scheduled for 2020, with improved astrometric solutions and completeness for the high proper motion stars, will enable more predictions much further
into the future.


\begin{acknowledgements}

I acknowledge the support of the NYU Abu Dhabi Research Enhancement Fund under grant RE124.
I thank the referee for taking the time to read this paper in detail and make insightful comments.
Many of the calculations performed in this paper employed code from the {\tt DanIDL} library of {\tt IDL} routines (\citealt{bra2017}) available
at \url{http://www.danidl.co.uk}. Martin Dominik gave very useful advice on references for the introduction and 
Lasha Ephremidze kindly checked some of my mathematical derivations.
I am grateful to Pierre Bergeron for providing the white dwarf cooling models in the \textit{Gaia} passbands.
This research was carried out on the High Performance Computing resources at New York University Abu Dhabi.
Thanks goes to Nasser Al Ansari, Muataz Al Barwani, Guowei He and Fayizal Mohammed Kunhi for their excellent
support. I would also like to thank the Solar System Dynamics Group at the Jet Propulsion Laboratory
for providing the Horizons On-Line Ephemeris System.
I made extensive use of the SIMBAD and VizieR web-resources as provided by
the Centre de Donn\'ees astronomiques de Strasbourg, for which I am exceptionally grateful.
This work has made use of data from the European Space Agency (ESA) mission
\textit{Gaia} (\url{https://www.cosmos.esa.int/gaia}), processed by the \textit{Gaia}
Data Processing and Analysis Consortium (DPAC,
\url{https://www.cosmos.esa.int/web/gaia/dpac/consortium}). Funding for the DPAC
has been provided by national institutions, in particular the institutions
participating in the \textit{Gaia} Multilateral Agreement.

Various people have encouraged and inspired me during 
my work on this topic. Thanks goes to
Irene Skuballa, Martin Bo Nielsen, Jasmina Blecic, Yanping Chen, Dave Russell, Mallory Roberts, Andrea Macci\`o, and Rocio and Lucia Mu\~niz Santacoloma.
I dedicate this work to my most wonderful daughters Phoebe and Chloe Bramich
Mu\~niz. Without their patience and support, this paper would have taken much longer to write.

\end{acknowledgements}



\begin{landscape}
\begin{table}
\centering
\caption{Characteristics of the nine microlensing events ME1-ME9 caused by the white dwarf lens star LAWD~37. Most quantities have already been defined in the text.
         $\Delta\theta_{2}$ is the difference between the minimum and maximum values of $\theta_{2}$ for an event over the 12-year baseline of an extended \textit{Gaia} mission.
         $T[\Delta\theta_{2}]$ is the amount of time that an event spends with $\theta_{2} > \min \{ \theta_{2} \} + \Delta\theta_{2}/2$.
         The numbers in parentheses indicate the uncertainty on the last digit.}
\small{
\begin{tabular}{@{}l|c|ccccc}
\hline
Name                                            & LAWD 37                    & ME1                         & ME2                        & ME3                         & ME4                        & ME5                        \\
\hline
GDR2 Source ID                                  & 5332606522595645952        & 5332606556930416896         & 5332606552624953344        & 5332606346466523008         & 5332606350796955904        & 5332606350775722752        \\
$\alpha_{\mbox{\scriptsize ref}}$ (deg$\pm$mas) & 176.4557771399$\pm$0.031   & 176.4523022357$\pm$0.399    & 176.4538556773$\pm$0.118   & 176.4594170339$\pm$0.103    & 176.4636014416$\pm$0.206   & 176.4677633522$\pm$2.252   \\
$\delta_{\mbox{\scriptsize ref}}$ (deg$\pm$mas) & $-$64.8430049999$\pm$0.033 & $-$64.8407666948$\pm$0.358  & $-$64.8408845446$\pm$0.115 & $-$64.8411595678$\pm$0.107  & $-$64.8432980750$\pm$0.200 & $-$64.8420809356$\pm$2.134 \\
$\mu_{\alpha*}$ (mas/year)                      & 2661.594(57)               & $-$4.88(103)                & $-$5.30(21)                & $-$7.08(19)                 & $-$8.62(37)                & -                          \\
$\mu_{\delta}$ (mas/year)                       & $-$344.847(59)             & 0.08(71)                    & 1.90(20)                   & $-$0.39(19)                 & 0.35(35)                   & -                          \\
$\varpi$ (mas)                                  & 215.766(41)                & $-$0.64(49)                 & $-$0.06(13)                & 0.19(12)                    & 0.42(25)                   & -                          \\
$G$ (mag)                                       & 11.4318(4)                 & 19.657(5)                   & 17.7332(11)                & 17.5213(10)                 & 18.6055(22)                & 20.946(14)                 \\
$G_{\mbox{\scriptsize BP}}$ (mag)               & 11.5072(43)                & 20.437(185)                 & 18.9103(551)               & 18.3276(274)                & 19.1711(603)               & 20.806(344)                \\
$G_{\mbox{\scriptsize RP}}$ (mag)               & 11.2352(9)                 & 18.603(58)                  & 16.5118(75)                & 16.5718(88)                 & 17.5823(226)               & 19.237(111)                \\
$M$ ($M_{\sun}$)                                & 0.78                       & -                           & -                          & -                           & -                          & -                          \\
$\theta_{\mbox{\scriptsize E}}$ (mas)           & -                          & 37.06(4)                    & 37.01(1)                   & 36.99(1)                    & 36.97(2)                   & 37.01(1)                   \\
$u_{0}$ ($\theta_{\mbox{\scriptsize E}}$)       & -                          & 217.50(25)                  & 193.97(7)                  & 194.71(6)                   & 9.95(5)                    & 149.43(7)                  \\
$u_{0}$ (mas)                                   & -                          & 8061.4(9)                   & 7179.4(3)                  & 7202.9(3)                   & 368.1(16)                  & 5530.3(24)                 \\
$t_{0}$ (Julian year)                           & -                          & 2014.56126$^{\mathrm{(a)}}$ & 2014.62854(7)              & 2016.81259(7)               & 2019.86776(34)             & 2021.80698(14)             \\
$\Delta\theta_{2}$ (mas)                        & -                          & 0.133(2)                    & 0.151(2)                   & 0.137(2)                    & 3.600(18)                  & 0.183(2)                   \\
$T[\Delta\theta_{2}]$ (d)                       & -                          & 955.3(5)                    & 1127.5(1)                  & 2176.8(2)                   & 122.2(6)                   & 1852.3(4)                  \\
\hline
Name                                            & LAWD 37                    & ME6                         & ME7                        & ME8                         & ME9                         \\
\hline
GDR2 Source ID                                  & 5332606522595645952        & 5332606350796954240         & 5332606346480229376        & 5332606350771989376         & 5332606350774673536         \\
$\alpha_{\mbox{\scriptsize ref}}$ (deg$\pm$mas) & 176.4557771399$\pm$0.031   & 176.4683954533$\pm$0.192    & 176.4697738660$\pm$0.237   & 176.4778826854$\pm$0.239    & 176.4787869637$\pm$1.680    \\
$\delta_{\mbox{\scriptsize ref}}$ (deg$\pm$mas) & $-$64.8430049999$\pm$0.033 & $-$64.8452675511$\pm$0.208  & $-$64.8428841195$\pm$0.205 & $-$64.8454278721$\pm$0.225  & $-$64.8439211575$\pm$1.690  \\
$\mu_{\alpha*}$ (mas/year)                      & 2661.594(57)               & $-$6.40(35)                 & $-$4.50(42)                & $-$4.99(49)                 & -                           \\
$\mu_{\delta}$ (mas/year)                       & $-$344.847(59)             & 1.14(35)                    & 1.92(35)                   & 1.30(46)                    & -                           \\
$\varpi$ (mas)                                  & 215.766(41)                & 0.35(23)                    & 0.12(26)                   & 0.11(26)                    & -                           \\
$G$ (mag)                                       & 11.4318(4)                 & 18.4416(21)                 & 18.7246(23)                & 18.8851(26)                 & 20.810(17)                  \\
$G_{\mbox{\scriptsize BP}}$ (mag)               & 11.5072(43)                & -                           & 19.3772(696)               & 19.5095(709)                & 20.283(129)                 \\
$G_{\mbox{\scriptsize RP}}$ (mag)               & 11.2352(9)                 & -                           & 17.7848(156)               & 17.9280(191)                & 19.530(124)                 \\
$M$ ($M_{\sun}$)                                & 0.78                       & -                           & -                          & -                           & -                           \\
$\theta_{\mbox{\scriptsize E}}$ (mas)           & -                          & 36.98(2)                    & 37.00(2)                   & 37.00(2)                    & 37.01(1)                    \\
$u_{0}$ ($\theta_{\mbox{\scriptsize E}}$)       & -                          & 146.81(9)                   & 83.49(10)                  & 182.21(17)                  & 161.22(5)                   \\
$u_{0}$ (mas)                                   & -                          & 5428.7(26)                  & 3089.2(29)                 & 6741.8(53)                  & 5966.4(18)                  \\
$t_{0}$ (Julian year)                           & -                          & 2023.21348(8)               & 2023.63689(62)             & 2026.56126$^{\mathrm{(b)}}$ & 2026.56126$^{\mathrm{(b)}}$ \\
$\Delta\theta_{2}$ (mas)                        & -                          & 0.194(2)                    & 0.386(4)                   & 0.167(3)                    & 0.194(2)                    \\
$T[\Delta\theta_{2}]$ (d)                       & -                          & 1830.4(6)                   & 1193.4(7)                  & 738.8(5)                    & 639.4(1)                    \\
\hline
\end{tabular}
}
\tablefoot{(a) Event peaks before this date and the event properties are only computed using the time period considered in this paper.
           (b) Event peaks after this date. For event properties computed from a more appropriate time period, see \citet{bra2018}.}
\label{tab:lawd37}
\end{table}
\end{landscape}

\afterpage{\clearpage}

\begin{landscape}
\begin{table}
\centering
\caption{Characteristics of the five microlensing events ME10-ME14 caused by the white dwarf lens star Stein~2051~B. The quantities are the same as in Table~\ref{tab:lawd37}.
         The numbers in parentheses indicate the uncertainty on the last digit.}
\small{
\begin{tabular}{@{}l|c|ccccc}
\hline
Name                                            & Stein 2051 B            & ME10                        & ME11                    & ME12                    & ME13                    & ME14                        \\
\hline
GDR2 Source ID                                  & 470826482635701376      & 470826482630688640          & 470826478336759552      & 470826478336759424      & 470826482633793664      & 470826478336745728          \\
$\alpha_{\mbox{\scriptsize ref}}$ (deg$\pm$mas) & 67.8135323847$\pm$0.056 & 67.8125405010$\pm$1.685     & 67.8178337100$\pm$0.315 & 67.8210828339$\pm$0.244 & 67.8217358439$\pm$9.297 & 67.8229392631$\pm$0.566     \\
$\delta_{\mbox{\scriptsize ref}}$ (deg$\pm$mas) & 58.9697516898$\pm$0.055 & 58.9704725133$\pm$1.085     & 58.9652134387$\pm$0.235 & 58.9657730960$\pm$0.185 & 58.9653011703$\pm$2.142 & 58.9620242635$\pm$0.463     \\
$\mu_{\alpha*}$ (mas/year)                      & 1335.042(98)            & 0.1(24)                     & 0.92(48)                & 2.50(37)                & -                       & $-$1.77(97)                 \\
$\mu_{\delta}$ (mas/year)                       & $-$1947.632(113)        & $-$0.3(20)                  & $-$3.12(49)             & $-$3.75(35)             & -                       & 0.21(90)                    \\
$\varpi$ (mas)                                  & 181.311(56)             & $-$0.1(14)                  & 0.73(37)                & 0.18(27)                & -                       & 0.74(66)                    \\
$G$ (mag)                                       & 12.3527(5)              & 19.734(14)                  & 18.7790(29)             & 18.5055(25)             & 20.949(22)              & 19.8477(59)                 \\
$G_{\mbox{\scriptsize BP}}$ (mag)               & 12.4942(44)             & 19.519(364)                 & 19.4761(377)            & 19.1929(333)            & -                       & 20.4126(1081)               \\
$G_{\mbox{\scriptsize RP}}$ (mag)               & 12.0602(14)             & 18.474(109)                 & 17.8765(188)            & 17.5917(283)            & -                       & 19.0055(453)                \\
$M$ ($M_{\sun}$)                                & 0.79                    & -                           & -                       & -                       & -                       & -                           \\
$\theta_{\mbox{\scriptsize E}}$ (mas)           & -                       & 34.09(13)                   & 34.01(4)                & 34.07(3)                & 34.08(1)                & 34.01(6)                    \\
$u_{0}$ ($\theta_{\mbox{\scriptsize E}}$)       & -                       & 33.23(10)                   & 73.34(15)               & 96.80(13)               & 97.59(25)               & 196.24(43)                  \\
$u_{0}$ (mas)                                   & -                       & 1132.6(39)                  & 2494.3(38)              & 3297.4(30)              & 3326.0(84)              & 6674.6(104)                 \\
$t_{0}$ (Julian year)                           & -                       & 2014.56126$^{\mathrm{(a)}}$ & 2023.21232(40)          & 2023.76194(22)          & 2024.74483(27)          & 2026.56126$^{\mathrm{(b)}}$ \\
$\Delta\theta_{2}$ (mas)                        & -                       & 0.986(6)                    & 0.406(2)                & 0.299(1)                & 0.301(1)                & 0.140(1)                    \\
$T[\Delta\theta_{2}]$ (d)                       & -                       & 207.0(7)                    & 1082.0(22)              & 1421.5(15)              & 1421.3(13)              & 703.4(14)                   \\
\hline
\end{tabular}
}
\tablefoot{(a) Event peaks before this date and the event properties are only computed using the time period considered in this paper.
           (b) Event peaks after this date. For event properties computed from a more appropriate time period, see \citet{bra2018}.}
\label{tab:stein2051b}
\end{table}
\end{landscape}

\afterpage{\clearpage}

\begin{landscape}
\begin{table}
\centering
\caption{Characteristics of the six photometric microlensing events ME15-ME20. Most quantities have already been defined in the text.
         $\Delta(A,A_{1})$ is the difference between the minimum and maximum magnifications (in magnitudes) of an event over the
         12-year baseline of an extended \textit{Gaia} mission.
         $T[\Delta(A,A_{1})]$ is the amount of time that an event spends with its magnification above
         $\min \{ A,A_{1} \} + \Delta(A,A_{1})/2$.
         Similarly, $\Delta(\delta_{\mbox{\scriptsize mic}},\theta_{2})$ is the difference between the minimum and maximum astrometric shifts
         of an event over the 12-year baseline of an extended \textit{Gaia} mission.
         $T[\Delta(\delta_{\mbox{\scriptsize mic}},\theta_{2})]$ is the amount of time that an event spends with its astrometric shift above
         $\min \{ \delta_{\mbox{\scriptsize mic}},\theta_{2} \} + \Delta(\delta_{\mbox{\scriptsize mic}},\theta_{2})/2$.
         The numbers in parentheses indicate the uncertainty on the last digit.}
\small{
\begin{tabular}{@{}l|cc|cc|cc}
\hline
Name                                            & OGLE SMC115.5.319          & ME15                            & SDSS J035037.54+112707.9 & ME16                         & -                       & ME17                         \\
\hline
Spectral Type                                   & WD                         & -                               & M2V                      & -                            & M                       & -                            \\
GDR2 Source ID                                  & 4687445500635789184        & 4687445599404851456             & 36697403171648128        & 36697407465577856            & 45148627499075584       & 45148631792418560            \\
$\alpha_{\mbox{\scriptsize ref}}$ (deg$\pm$mas) & 17.7691778251$\pm$0.123    & 17.7707589355$\pm$0.169         & 57.6566688599$\pm$0.234  & 57.6568631007$\pm$0.856      & 61.8675521451$\pm$0.268 & 61.8676334642$\pm$1.373      \\
$\delta_{\mbox{\scriptsize ref}}$ (deg$\pm$mas) & $-$72.4621946776$\pm$0.090 & $-$72.4624613402$\pm$0.122      & 11.4519500665$\pm$0.110  & 11.4517550481$\pm$0.757      & 15.3066735721$\pm$0.186 & 15.3065150112$\pm$1.267      \\
$\mu_{\alpha*}$ (mas/year)                      & 272.35(24)                 & 1.02(34)                        & 91.89(45)                & -                            & 34.94(62)               & -                            \\
$\mu_{\delta}$ (mas/year)                       & $-$170.39(18)              & $-$1.78(24)                     & $-$97.02(31)             & -                            & $-$67.03(53)            & -                            \\
$\varpi$ (mas)                                  & 34.84(12)                  & 0.10(17)                        & 2.74(27)                 & -                            & 3.71(29)                & -                            \\
$G$ (mag)                                       & 17.6482(12)                & 18.1510(18)                     & 17.8776(13)              & 19.4261(41)                  & 18.5799(24)             & 20.347(10)                   \\
$G_{\mbox{\scriptsize BP}}$ (mag)               & 18.0831(434)               & 17.9983(89)                     & 18.7822(212)             & -                            & 19.7360(667)            & -                            \\
$G_{\mbox{\scriptsize RP}}$ (mag)               & 16.8239(107)               & 18.0555(34)                     & 16.6000(56)              & -                            & 17.1440(122)            & -                            \\
$M$ ($M_{\sun}$)                                & 0.64                       & -                               & 0.46                     & -                            & 0.42                    & -                            \\
$\theta_{\mbox{\scriptsize E}}$ (mas)           & -                          & 13.42(4)                        & -                        & 3.21(16)                     & -                       & 3.58(14)                     \\
$u_{0}$ ($\theta_{\mbox{\scriptsize E}}$)       & -                          & 5.25$^{+0.16}_{-0.15}$          & -                        & 4.32$^{+1.15}_{-1.01}$       & -                       & 3.47$^{+1.60}_{-1.58}$       \\
$u_{0}$ (mas)                                   & -                          & 70.43$^{+2.18}_{-2.06}$         & -                        & 13.90$^{+3.34}_{-3.15}$      & -                       & 12.43$^{+5.57}_{-5.72}$      \\
$t_{0}$ (Julian year)                           & -                          & 2021.4941$^{+0.0032}_{-0.0029}$ & -                        & 2022.831(20)                 & -                       & 2023.987$^{+0.050}_{-0.054}$ \\
$\Delta(A,A_{1})$ (mag)                         & -                          & 0.0010(1)                       & -                        & 0.0010$^{+0.0016}_{-0.0006}$ & -                       & 0.0019$^{+0.0110}_{-0.0014}$ \\
$T[\Delta(A,A_{1})]$ (d)                        & -                          & 62.5$^{+1.8}_{-1.7}$            & -                        & 54.8$^{+11.6}_{-10.9}$       & -                       & 80.9$^{+29.8}_{-29.7}$       \\
$\Delta(\delta_{\mbox{\scriptsize mic}},\theta_{2})$ (mas) & -               & 1.638$^{+0.009}_{-0.010}$       & -                        & 0.132$^{+0.042}_{-0.031}$    & -                       & 0.148$^{+0.111}_{-0.039}$    \\
$T[\Delta(\delta_{\mbox{\scriptsize mic}},\theta_{2})]$ (d) & -              & 339.2$^{+4.1}_{-6.1}$           & -                        & 333$^{+195}_{-189}$          & -                       & 530$^{+419}_{-409}$          \\
\hline
Name                                            & G192-52                  & ME18                         & 2MASS J13055171$-$7218081  & ME19                            & -                          & ME20                            \\
\hline
Spectral Type                                   & sdM1e                    & -                            & M                          & -                               & WD                         & -                               \\
GDR2 Source ID                                  & 993144190006664832       & 993144185711067136           & 5840411363658156032        & 5840411359350016128             & 5902615306303778304        & 5902615301994733696             \\
$\alpha_{\mbox{\scriptsize ref}}$ (deg$\pm$mas) & 102.6904889155$\pm$0.031 & 102.6906795987$\pm$0.684     & 196.4603985004$\pm$0.087   & 196.4590182889$\pm$0.159        & 229.5240609423$\pm$0.095   & 229.5240237978$\pm$0.104        \\
$\delta_{\mbox{\scriptsize ref}}$ (deg$\pm$mas) & 52.2282352111$\pm$0.028  & 52.2272715905$\pm$0.658      & $-$72.3009951372$\pm$0.076 & $-$72.3006261030$\pm$0.136      & $-$48.9059371210$\pm$0.079 & $-$48.9063514134$\pm$0.087      \\
$\mu_{\alpha*}$ (mas/year)                      & 39.673(52)               & 0.5(10)                      & $-$362.72(17)              & $-$12.95(37)                    & $-$18.15(26)               & $-$3.85(29)                     \\
$\mu_{\delta}$ (mas/year)                       & $-$337.940(52)           & $-$1.2(11)                   & 306.51(15)                 & 2.93(29)                        & $-$159.22(22)              & $-$3.28(24)                     \\
$\varpi$ (mas)                                  & 13.394(38)               & $-$0.2(7)                    & 9.55(9)                    & 0.62(17)                        & 24.64(13)                  & 0.19(14)                        \\
$G$ (mag)                                       & 13.6801(4)               & 19.5038(62)                  & 17.1820(8)                 & 18.1750(17)                     & 17.5120(13)                & 17.6448(12)                     \\
$G_{\mbox{\scriptsize BP}}$ (mag)               & 14.7009(32)              & 19.1053(2084)                & 18.1869(276)               & 18.7933(613)                    & 17.5671(279)               & -                               \\
$G_{\mbox{\scriptsize RP}}$ (mag)               & 12.6886(15)              & 17.7353(687)                 & 16.0422(117)               & 17.3676(479)                    & 16.5060(430)               & -                               \\
$M$ ($M_{\sun}$)                                & 0.47                     & -                            & 0.25                       & -                               & 0.45                       & -                               \\
$\theta_{\mbox{\scriptsize E}}$ (mas)           & -                        & 7.19(19)                     & -                          & 4.25(5)                         & -                          & 9.51(4)                         \\
$u_{0}$ ($\theta_{\mbox{\scriptsize E}}$)       & -                        & 1.16$^{+1.11}_{-0.79}$       & -                          & 1.16$^{+0.36}_{-0.37}$          & -                          & 2.69$^{+0.35}_{-0.39}$          \\
$u_{0}$ (mas)                                   & -                        & 8.31$^{+8.17}_{-5.63}$       & -                          & 4.93$^{+1.49}_{-1.56}$          & -                          & 25.51$^{+3.35}_{-3.66}$         \\
$t_{0}$ (Julian year)                           & -                        & 2025.810$^{+0.040}_{-0.038}$ & -                          & 2019.8388$^{+0.0036}_{-0.0033}$ & -                          & 2025.0843$^{+0.0086}_{-0.0083}$ \\
$\Delta(A,A_{1})$ (mag)                         & -                        & 0.0013$^{+0.0080}_{-0.0011}$ & -                          & 0.074$^{+0.084}_{-0.035}$       & -                          & 0.0124$^{+0.0078}_{-0.0041}$    \\
$T[\Delta(A,A_{1})]$ (d)                        & -                        & 19.4$^{+11.8}_{-10.7}$       & -                          & 7.68$^{+1.58}_{-1.75}$          & -                          & 42.6$^{+3.9}_{-4.4}$            \\
$\Delta(\delta_{\mbox{\scriptsize mic}},\theta_{2})$ (mas) & -             & 0.498(26)                    & -                          & 0.718(93)                       & -                          & 1.415$^{+0.182}_{-0.132}$       \\
$T[\Delta(\delta_{\mbox{\scriptsize mic}},\theta_{2})]$ (d) & -            & 191.7(26)                    & -                          & 22.0$^{+2.8}_{-2.0}$            & -                          & 200$^{+158}_{-86}$              \\
\hline
\end{tabular}
}
\label{tab:photevents1}
\end{table}
\end{landscape}

\afterpage{\clearpage}

\begin{landscape}
\begin{table}
\centering
\caption{Characteristics of the three photometric microlensing events ME21-ME23. The quantities are the same as in Table~\ref{tab:photevents1}.
         The numbers in parentheses indicate the uncertainty on the last digit.}
\small{
\begin{tabular}{@{}l|cc|cc|cc}
\hline
Name                                            & -                          & ME21                         & Wolf 851                 & ME22                            & -                        & ME23                            \\
\hline
Spectral Type                                   & WD                         & -                            & sdK7                     & -                               & M                        & -                               \\
GDR2 Source ID                                  & 4079513805001070720        & 4079513805001070592          & 4278722497040124032      & 4278722497031268224             & 2026434843801289344      & 2026434843801289472             \\
$\alpha_{\mbox{\scriptsize ref}}$ (deg$\pm$mas) & 278.7987836452$\pm$0.253   & 278.7985968980$\pm$28.409    & 280.4016967890$\pm$0.050 & 280.4018007595$\pm$1.142        & 291.2523104521$\pm$0.118 & 291.2522150836$\pm$0.122        \\
$\delta_{\mbox{\scriptsize ref}}$ (deg$\pm$mas) & $-$21.9809041462$\pm$0.246 & $-$21.9807651294$\pm$28.781  & 0.9119814724$\pm$0.046   & 0.9075026888$\pm$1.059          & 28.7451149554$\pm$0.142  & 28.7448781489$\pm$0.139         \\
$\mu_{\alpha*}$ (mas/year)                      & $-$73.10(51)               & -                            & 45.95(11)                & $-$3.0(24)                      & $-$31.73(27)             & $-$0.57(27)                     \\
$\mu_{\delta}$ (mas/year)                       & 55.47(43)                  & -                            & $-$1981.23(10)           & $-$4.6(24)                      & $-$81.80(31)             & $-$2.24(29)                     \\
$\varpi$ (mas)                                  & 12.41(28)                  & -                            & 33.325(51)               & 0.3(13)                         & 5.70(17)                 & $-$0.30(20)                     \\
$G$ (mag)                                       & 18.1970(45)                & 19.9788(98)                  & 11.5616(3)               & 20.415(14)                      & 18.2119(17)              & 18.1091(19)                     \\
$G_{\mbox{\scriptsize BP}}$ (mag)               & 18.3191(491)               & -                            & 12.4457(19)              & -                               & 18.7117(861)             & 18.4817(217)                    \\
$G_{\mbox{\scriptsize RP}}$ (mag)               & 17.4404(651)               & -                            & 10.6595(18)              & 19.107(94)                      & 16.4903(312)             & 16.4346(136)                    \\
$M$ ($M_{\sun}$)                                & 0.32                       & -                            & 0.43                     & -                               & 0.24                     & -                               \\
$\theta_{\mbox{\scriptsize E}}$ (mas)           & -                          & 5.71(7)                      & -                        & 10.69(21)                       & -                        & 3.45(8)                         \\
$u_{0}$ ($\theta_{\mbox{\scriptsize E}}$)       & -                          & 4.81$^{+5.26}_{-3.40}$       & -                        & 1.22$^{+1.46}_{-0.86}$          & -                        & 7.80$^{+1.19}_{-1.11}$          \\
$u_{0}$ (mas)                                   & -                          & 27.4$^{+30.2}_{-19.4}$       & -                        & 13.1$^{+15.0}_{-9.4}$           & -                        & 27.0$^{+4.2}_{-4.0}$            \\
$t_{0}$ (Julian year)                           & -                          & 2024.266$^{+0.166}_{-0.083}$ & -                        & 2023.6559$^{+0.0107}_{-0.0097}$ & -                        & 2026.1603$^{+0.0299}_{-0.0327}$ \\
$\Delta(A,A_{1})$ (mag)                         & -                          & 0.0006$^{+0.0264}_{-0.0005}$ & -                        & 0.0001$^{+0.0005}_{-0.0001}$    & -                        & 0.0003$^{+0.0002}_{-0.0001}$    \\
$T[\Delta(A,A_{1})]$ (d)                        & -                          & 121$^{+86}_{-80}$            & -                        & 7.45$^{+0.78}_{-5.75}$          & -                        & 133(15)                         \\
$\Delta(\delta_{\mbox{\scriptsize mic}},\theta_{2})$ (mas) & -               & 0.282$^{+0.217}_{-0.009}$    & -                        & 1.10(4)                         & -                        & 0.216$^{+0.038}_{-0.029}$       \\
$T[\Delta(\delta_{\mbox{\scriptsize mic}},\theta_{2})]$ (d) & -              & 699$^{+129}_{-450}$          & -                        & 38.78$^{+0.58}_{-0.18}$         & -                        & 336$^{+72}_{-41}$               \\
\hline
\end{tabular}
}
\label{tab:photevents2}
\end{table}
\end{landscape}

\afterpage{\clearpage}

\begin{landscape}
\begin{table}
\centering
\caption{Characteristics of the six astrometric microlensing events ME24-ME29. The quantities are the same as in Table~\ref{tab:photevents1}.
         The numbers in parentheses indicate the uncertainty on the last digit.}
\small{
\begin{tabular}{@{}l|cc|cc|cc}
\hline
Name                                            & LSPM J0024+6834N        & ME24                    & Ross 322                & ME25                    & L~51-47                    & ME26                       \\
\hline
Spectral Type                                   & WD                      & -                       & M3V                     & -                       & M3-4V                      & -                          \\
GDR2 Source ID                                  & 529594417061837824      & 529594417064722944      & 314922605759778048      & 314922601464808064      & 4687511776265158400        & 4687511780573305984        \\
$\alpha_{\mbox{\scriptsize ref}}$ (deg$\pm$mas) & 6.2147084027$\pm$0.091  & 6.2176397228$\pm$2.737  & 16.9564920313$\pm$0.057 & 16.9579186795$\pm$0.331 & 17.3299573806$\pm$0.038    & 17.3312978748$\pm$0.044    \\
$\delta_{\mbox{\scriptsize ref}}$ (deg$\pm$mas) & 68.5797798551$\pm$0.075 & 68.5799164329$\pm$1.636 & 34.2105534163$\pm$0.046 & 34.2110043370$\pm$0.476 & $-$72.2098406097$\pm$0.030 & $-$72.2094955434$\pm$0.036 \\
$\mu_{\alpha*}$ (mas/year)                      & 411.63(21)              & -                       & 1373.673(107)           & $-$1.12(75)             & 583.083(92)                & $-$28.303(117)             \\
$\mu_{\delta}$ (mas/year)                       & 23.07(13)               & -                       & 480.333(98)             & 0.83(77)                & 439.494(60)                & $-$20.443(84)              \\
$\varpi$ (mas)                                  & 27.70(10)               & -                       & 42.538(61)              & $-$0.02(56)             & 51.931(36)                 & 3.695(44)                  \\
$G$ (mag)                                       & 17.4786(10)             & 20.821(14)              & 12.3722(4)              & 18.6144(51)             & 11.8735(4)                 & 14.8967(25)                \\
$G_{\mbox{\scriptsize BP}}$ (mag)               & 17.7780(91)             & 20.178(602)             & 13.6244(31)             & 18.5805(1051)           & 13.1591(53)                & -                          \\
$G_{\mbox{\scriptsize RP}}$ (mag)               & 16.8762(131)            & 18.854(38)              & 11.2706(9)              & 17.5584(960)            & 10.7492(30)                & -                          \\
$M$ ($M_{\sun}$)                                & 0.68                    & -                       & 0.38                    & -                       & 0.33                       & -                          \\
$\theta_{\mbox{\scriptsize E}}$ (mas)           & -                       & 12.407(23)              & -                       & 11.430(78)              & -                          & 11.418(7)                  \\
$u_{0}$ ($\theta_{\mbox{\scriptsize E}}$)       & -                       & 20.61(17)               & -                       & 11.86(18)               & -                          & 6.03(3)                    \\
$u_{0}$ (mas)                                   & -                       & 255.7(20)               & -                       & 135.5(21)               & -                          & 68.89(33)                  \\
$t_{0}$ (Julian year)                           & -                       & 2024.7992(29)           & -                       & 2018.6059(19)           & -                          & 2018.06742(25)             \\
$\Delta\theta_{2}$ (mas)                        & -                       & 0.564(6)                & -                       & 0.945(18)               & -                          & 1.230(1)                   \\
$T[\Delta\theta_{2}]$ (d)                       & -                       & 706.7(57)               & -                       & 124.0(20)               & -                          & 87.33(10)                  \\
\hline
Name                                            & -                          & ME27                       & -                          & ME28                       & Ross~528                   & ME29                       \\
\hline
Spectral Type                                   & WD                         & -                          & M                          & -                          & K4V                        & -                          \\
GDR2 Source ID                                  & 5355886688435657344        & 5355886688435657216        & 5239052307989553664        & 5239052303673076224        & 6246397442267602432        & 6246397437969967872        \\
$\alpha_{\mbox{\scriptsize ref}}$ (deg$\pm$mas) & 155.2385973786$\pm$0.194   & 155.2381806130$\pm$0.482   & 160.9623938494$\pm$0.224   & 160.9620833874$\pm$0.060   & 245.1676860798$\pm$0.078   & 245.1666624383$\pm$0.309   \\
$\delta_{\mbox{\scriptsize ref}}$ (deg$\pm$mas) & $-$53.1229462715$\pm$0.170 & $-$53.1223436683$\pm$0.485 & $-$66.2926856760$\pm$0.186 & $-$66.2923895767$\pm$0.060 & $-$17.6516704757$\pm$0.037 & $-$17.6529768694$\pm$0.175 \\
$\mu_{\alpha*}$ (mas/year)                      & $-$109.47(39)              & $-$7.0(12)                 & $-$39.49(41)               & $-$6.89(13)                & $-$319.87(14)              & $-$0.76(60)                \\
$\mu_{\delta}$ (mas/year)                       & 193.75(36)                 & 3.2(12)                    & 118.69(33)                 & 11.53(11)                  & $-$468.19(10)              & $-$16.23(44)               \\
$\varpi$ (mas)                                  & 23.62(20)                  & 0.88(56)                   & 24.58(22)                  & 1.091(70)                  & 20.888(85)                 & 0.67(35)                   \\
$G$ (mag)                                       & 18.4970(20)                & 19.6327(53)                & 18.1999(27)                & 16.4419(8)                 & 10.8750(8)                 & 18.2841(60)                \\
$G_{\mbox{\scriptsize BP}}$ (mag)               & 18.9915(286)               & -                          & -                          & 16.8648(210)               & 11.6155(11)                & 18.4332(1041)              \\
$G_{\mbox{\scriptsize RP}}$ (mag)               & 17.8582(184)               & -                          & -                          & 15.5586(113)               & 10.0686(9)                 & 16.9804(1297)              \\
$M$ ($M_{\sun}$)                                & 0.78                       & -                          & 0.46                       & -                          & 0.65                       & -                          \\
$\theta_{\mbox{\scriptsize E}}$ (mas)           & -                          & 12.01(16)                  & -                          & 9.416(45)                  & -                          & 10.366(92)                 \\
$u_{0}$ ($\theta_{\mbox{\scriptsize E}}$)       & -                          & 19.61$^{+0.97}_{-0.92}$    & -                          & 10.77(45)                  & -                          & 16.49$^{+0.67}_{-0.63}$    \\
$u_{0}$ (mas)                                   & -                          & 235.5$^{+10.6}_{-9.7}$     & -                          & 101.4(42)                  & -                          & 170.9$^{+6.8}_{-6.3}$      \\
$t_{0}$ (Julian year)                           & -                          & 2025.900(14)               & -                          & 2025.4444(60)              & -                          & 2026.097(12)               \\
$\Delta\theta_{2}$ (mas)                        & -                          & 0.553$^{+0.034}_{-0.031}$  & -                          & 0.778$^{+0.010}_{-0.017}$  & -                          & 0.609(28)                  \\
$T[\Delta\theta_{2}]$ (d)                       & -                          & 733$^{+16}_{-18}$          & -                          & 815$^{+5}_{-4}$            & -                          & 373(6)                     \\
\hline
\end{tabular}
}
\label{tab:astromevents1}
\end{table}
\end{landscape}

\afterpage{\clearpage}

\begin{landscape}
\begin{table}
\centering
\caption{Characteristics of the four astrometric microlensing events ME30-ME33. The quantities are the same as in Table~\ref{tab:photevents1}.
         The numbers in parentheses indicate the uncertainty on the last digit.}
\small{
\begin{tabular}{@{}l|cc|cc|cc}
\hline
Name                                            & LAWD~66                    & ME30                            & WD~1743$-$545              & ME31                       & LSPM J1913+2949          & ME32                     \\
\hline
Spectral Type                                   & DQ6                        & -                               & DC                         & -                          & DA                       & -                        \\
GDR2 Source ID                                  & 4139531467491239680        & 4139531467491232000             & 5920900901901635968        & 5920900970620951424        & 2039140284770609152      & 2039140280460138112      \\
$\alpha_{\mbox{\scriptsize ref}}$ (deg$\pm$mas) & 257.8627461049$\pm$0.044   & 257.8634211454$\pm$15.072       & 266.9005413726$\pm$0.072   & 266.8984483271$\pm$0.133   & 288.3186969451$\pm$0.059 & 288.3185069118$\pm$0.251 \\
$\delta_{\mbox{\scriptsize ref}}$ (deg$\pm$mas) & $-$14.7994265360$\pm$0.035 & $-$14.8001630331$\pm$13.062     & $-$54.6099768183$\pm$0.065 & $-$54.6107190122$\pm$0.124 & 29.8257586483$\pm$0.072  & 29.8265686663$\pm$0.319  \\
$\mu_{\alpha*}$ (mas/year)                      & 278.156(115)               & -                               & $-$383.95(12)              & $-$1.63(22)                & $-$44.22(12)             & 0.26(52)                 \\
$\mu_{\delta}$ (mas/year)                       & $-$273.661(73)             & -                               & $-$306.68(10)              & $-$6.10(20)                & 283.65(15)               & $-$5.41(73)              \\
$\varpi$ (mas)                                  & 44.128(49)                 & -                               & 74.012(87)                 & 0.26(18)                   & 26.959(85)               & 0.07(35)                 \\
$G$ (mag)                                       & 14.2758(7)                 & 20.778(16)                      & 15.9489(8)                 & 17.3304(19)                & 16.9532(8)               & 19.3531(33)              \\
$G_{\mbox{\scriptsize BP}}$ (mag)               & 14.2606(45)                & -                               & 16.5473(111)               & 17.7175(155)               & 17.2237(91)              & 19.7860(1003)            \\
$G_{\mbox{\scriptsize RP}}$ (mag)               & 14.1797(20)                & -                               & 15.1990(17)                & 16.7546(60)                & 16.5113(42)              & 18.6461(232)             \\
$M$ ($M_{\sun}$)                                & 0.85                       & -                               & 0.54                       & -                          & 0.63                     & -                        \\
$\theta_{\mbox{\scriptsize E}}$ (mas)           & -                          & 17.495(10)                      & -                          & 18.068(25)                 & -                        & 11.787(78)               \\
$u_{0}$ ($\theta_{\mbox{\scriptsize E}}$)       & -                          & 12.0(11)                        & -                          & 29.51(16)                  & -                        & 10.32$^{+0.48}_{-0.53}$  \\
$u_{0}$ (mas)                                   & -                          & 210(20)                         & -                          & 533.3(25)                  & -                        & 121.7$^{+5.6}_{-6.1}$    \\
$t_{0}$ (Julian year)                           & -                          & 2024.6432$^{+0.0014}_{-0.0017}$ & -                          & 2025.84470(14)             & -                        & 2025.736(11)             \\
$\Delta\theta_{2}$ (mas)                        & -                          & 1.37(13)                        & -                          & 0.552(4)                   & -                        & 1.09(6)                  \\
$T[\Delta\theta_{2}]$ (d)                       & -                          & 532$^{+120}_{-57}$              & -                          & 781.2(14)                  & -                        & 419$^{+37}_{-32}$        \\
\hline
Name                                            & Ross 213                 & ME33                        \\
\hline
Spectral Type                                   & K5V                      & -                           \\
GDR2 Source ID                                  & 2215963091909478144      & 2215963121964404736         \\
$\alpha_{\mbox{\scriptsize ref}}$ (deg$\pm$mas) & 326.6875711035$\pm$0.026 & 326.6892675395$\pm$0.235    \\
$\delta_{\mbox{\scriptsize ref}}$ (deg$\pm$mas) & 62.0433818361$\pm$0.029  & 62.0439607150$\pm$0.223     \\
$\mu_{\alpha*}$ (mas/year)                      & 244.132(58)              & $-$3.91(41)                 \\
$\mu_{\delta}$ (mas/year)                       & 188.172(53)              & $-$3.96(35)                 \\
$\varpi$ (mas)                                  & 14.393(31)               & 0.46(32)                    \\
$G$ (mag)                                       & 11.7687(2)               & 17.8346(39)                 \\
$G_{\mbox{\scriptsize BP}}$ (mag)               & 12.5859(14)              & 17.7394(2284)               \\
$G_{\mbox{\scriptsize RP}}$ (mag)               & 10.8961(8)               & 16.3783(1578)               \\
$M$ ($M_{\sun}$)                                & 0.64                     & -                           \\
$\theta_{\mbox{\scriptsize E}}$ (mas)           & -                        & 8.52(10)                    \\
$u_{0}$ ($\theta_{\mbox{\scriptsize E}}$)       & -                        & 14.65$^{+0.51}_{-0.48}$     \\
$u_{0}$ (mas)                                   & -                        & 124.8$^{+5.2}_{-4.8}$       \\
$t_{0}$ (Julian year)                           & -                        & 2026.56126$^{\mathrm{(a)}}$ \\
$\Delta\theta_{2}$ (mas)                        & -                        & 0.560$^{+0.020}_{-0.019}$   \\
$T[\Delta\theta_{2}]$ (d)                       & -                        & 168.0$^{+6.5}_{-5.6}$       \\
\hline
\end{tabular}
}
\tablefoot{(a) Event peaks after this date. For event properties computed from a more appropriate time period, see \citet{bra2018}.}
\label{tab:astromevents2}
\end{table}
\end{landscape}

\afterpage{\clearpage}

\begin{landscape}
\begin{table}
\centering
\caption{Characteristics of the 43 astrometric microlensing events ME34-ME76. The quantities are the same as in Table~\ref{tab:photevents1}.
         The numbers in parentheses indicate the uncertainty on the last digit.}
\small{
\begin{tabular}{@{}l|cc}
\hline
Name & & \\
\hline
Spectral Type & & \\
GDR2 Source ID & & \\
$\alpha_{\mbox{\scriptsize ref}}$ (deg$\pm$mas) & & \\
$\delta_{\mbox{\scriptsize ref}}$ (deg$\pm$mas) & & \\
$\mu_{\alpha*}$ (mas/year) & & \\
$\mu_{\delta}$ (mas/year) & & \\
$\varpi$ (mas) & & \\
$G$ (mag) & & \\
$G_{\mbox{\scriptsize BP}}$ (mag) & & \\
$G_{\mbox{\scriptsize RP}}$ (mag) & & \\
$M$ ($M_{\sun}$) & & \\
$\theta_{\mbox{\scriptsize E}}$ (mas) & & \\
$u_{0}$ ($\theta_{\mbox{\scriptsize E}}$) & & \\
$u_{0}$ (mas) & & \\
$t_{0}$ (Julian year) & & \\
$\Delta\theta_{2}$ (mas) & & \\
$T[\Delta\theta_{2}]$ (d) & & \\
\hline
\end{tabular}
}
\label{tab:remainingevents}
\end{table}
\end{landscape}

\end{document}